\documentclass[useAMS,usenatbib]{mn2e}

\usepackage{amssymb}
\usepackage[fleqn]{amsmath}
\usepackage{graphicx}
\usepackage{latex_style_sf}
\usepackage{mdwlist}
\usepackage{float}
\usepackage{chngpage}
\usepackage{threeparttable}
\usepackage{subfigure}
\usepackage{needspace}

\makeatletter

\renewcommand{\@thesubfigure}{(\alph{subfigure})\hskip\subfiglabelskip}
\renewcommand{\@@thesubfigure}{(\alph{subfigure})}
\makeatother

\title[Eccentric Planets, Stellar Evolution, and Polluted WDs]{Eccentric Planets and Stellar Evolution as a Cause of Polluted White Dwarfs}
\author[S. Frewen \& B. Hansen] {S.~F.~N. Frewen,$^1$\thanks{E-mail: sfrewen@astro.ucla.edu} B.~M.~S. Hansen$^1$ \\
$^1$ Division of  Astronomy and Astrophysics, University of California, Los Angeles,  CA 90095-1547}

\pagerange{\pageref{firstpage}--\pageref{lastpage}} \pubyear{2013}

\begin{document}

\label{firstpage}

\maketitle

\begin{abstract}
A significant fraction of white dwarfs are observed to be polluted with metals despite high surface gravities and short settling times. The current theoretical model for this pollution is accretion of rocky bodies, which are delivered to the white dwarf through perturbations by orbiting planets. Using N--body simulations, we examine the possibility of a single planet as the source of pollution. We determine the stability of test particles on circular orbits in systems with a single planet located at 4 au for a range of masses and eccentricities, comparing the fractions that are ejected and accreted by the star. In particular, we compare the instabilities that develop before and after the star loses mass to form a white dwarf, a process which causes the semi-major axes of orbiting bodies to expand adiabatically. We determine that a planet must be eccentric ($e>0.02$) to deliver significant ($>0.5$ per cent) amounts of material to the central body, and that the amount increases with the planetary eccentricity. This result is robust with respect to the initial eccentricities of the scattered particles in the case of planetary eccentricity above $\sim0.4$ and the case of randomly-distributed particle longitude of pericentre.  We also find that the efficiency of the pollution is enhanced as planetary mass is reduced. We demonstrate that a 0.03 M$_\mathrm{Jup}$ planet with substantial eccentricity ($e>0.4$) can account for the observed levels of pollution for initial disc masses of order 1 M$_\oplus$. Such discs are well within the range estimated for initial planetesimals discs and well below that estimated for our own solar system within the context of the Nice model. However, their long term survival to the white dwarf stage is uncertain as estimates for the collisional evolution of planetesimal discs suggest they should be ground down below the required levels on Gyr timescales. Thus, planetary scattering by eccentric, sub-Jovian planets can explain the observed levels of pollution in white dwarfs, but only if current estimates of  the collisional erosion of planetesimal discs are in error.

\end{abstract}


\section{Introduction}\label{sec:intro}

Studies of hydrogen-dominated (DA) and helium-dominated (DB) white-dwarf (WD) spectra have found that roughly 25 per cent of all WDs show weak metal lines, despite the theoretical prediction that metals should gravitationally settle below the photosphere \citep{Zuck2003, Koester2009, Zuckerman:2010kl}. Given settling time-scales as small as days for the more-common DAs and under $10^5$ years for DBs, the rapid rate at which gravitational settling occurs in these dense objects indicates the pollution must have occurred recently compared to their cooling ages and is likely ongoing \citep{Paquette1986, Koester:2006kx, Koester2009}. Furthermore, infrared excesses have been detected in a number of these WD systems \citep{Zuckerman:1987ly,von-Hippel:2007yq,Farihi2009, Xu:2012ly}. The cause of these excesses appears to be debris discs near the WDs, which are delivering the polluting material \citep{Chary:1999mz,Jura2003,Kilic:2006vn}. These detections support the theory that the observed pollution is due to accreted orbiting bodies, which are perturbed towards and tidally disrupted by the central WD \citep{Jura2006}. The primary alternative, accretion of interstellar material, has been proven inconsistent with elemental abundances determined from  WD spectra \citep{Jura2006, Klein2010,Farihi:2010mz, Klein:2011cr} as well as spectra of the debris discs \citep{Reach:2005fr,Jura:2009ys}. Estimates for total mass accreted by polluted WDs are on the order of  $6\times10^{23}$ g, which is similar to the mass of minor bodies in the solar system such as Ceres \citep{Zuckerman:2010kl}.

While the source of the pollution has become more clear in recent years, the mechanism for delivering such material to the star is still uncertain. Early work investigated the possibility that planetary systems close to instability during the main sequence (MS) can be destabilized by stellar mass loss during post-MS evolution \citep{DebesSigurdsson}. Upon scattering, the planets could settle into a new, dynamically-young configuration that would allow them to perturb other orbiting bodies, such as asteroids or comets, that were previously stable. However, the high frequency of polluted WDs means this mechanism would require a large fraction of WD progenitors to have planetary systems on the edge of stability.

More recently, \citet{Bonsor2011} simulated planets on circular orbits interacting with a Kuiper-Belt analog over the course of stellar evolution. To determine the ability of the planet to pollute its host WD, the authors looked at the fraction of bodies scattered into the inner solar system, interior to the orbit of the planet. They did not simulate impacts with the WD; instead, they assumed additional planets in this region, which could further scatter bodies. The authors found that a planet near 30 au and a Kuiper Belt analog extending to 46.7 au could match the observed frequency of polluted WDs as well as the distribution of accretion rates as a function of cooling age. This result requires an inner planetary system to cause secondary scatterings, which delivers the material the full distance to the WD; a lone planet at 30 au was unable to scatter particles on to stellar-collision orbits. \citet{Bonsor2011} also found that the mass of the simulated planet had only a weak effect on the amount of material delivered to the inner system. While massive planets removed more bodies from the original belt, higher fractions were ejected relative to less-massive planets.

Another recent work, \citet{Debes2012}, focused on rocky bodies interior to a planet as source of pollution. The authors examined the ability of a single planet to deliver material originating in the inner 2:1 Mean Motion Resonance (MMR) to a WD, through the increase in resonance width with stellar evolution.  Using Jupiter and the asteroids located near the 2:1 MMR as a test case, they found that a single planet is capable of delivering enough material to the star provided the interior debris belt is large enough. Assuming a similar size distribution to the solar system asteroid belt, the authors determined that the total mass of the debris belt would need to be between 4 times and $6\times10^5$ times larger to account for WD observations. While massive debris discs have been observed, they often exist at greater distances from their host star \citep{Wyatt:2008th} and their incidence appears to drop off rapidly with age \citep{Rieke:2005fy}.

In a paper related to \citet{DebesSigurdsson}, \citet{Veras:2013vn} examined the stability of two-planet systems over the entirety of stellar evolution, including MS, post-MS (including mass loss), and WD. The authors performed N-body integrations of the system for stellar masses 3--8 M$_\odot$, with equal planetary masses of both M$_\mathrm{Jup}$ and M$_\oplus$. They found that planetary interactions leading to ejection or accretion by the star primarily occurred after $\sim10^7$ years, supporting the possibility of planetary instability as a source of WD pollution. However, in addition to selecting large stellar masses dissimilar to the progenitors of observed polluted WDs (1--2.5 M$_\odot$), the authors did not investigate the effect of the unstable planets on planetesimals remaining in the system, which limits the applicability of their results to WD pollution.

In this work we use N-body simulations of a single planet and massless test particles (TPs) to systematically examine the degree to which planetary properties affect WD pollution. In particular, we investigate the influence of planetary mass and eccentricity, which have largely been neglected in this context until now. Radial velocity surveys have shown that exoplanets orbiting beyond 0.1 au have a wide range of eccentricities \citep{Butler:2006yq}, and a cursory examination of the current Exoplanet Orbit Database shows that over over 50 per cent of such planets have eccentricities larger than 0.2 \citep{Wright:2011fk}. Therefore, testing a range of eccentricities for the perturbing planet gives us greater generality than before as well as insight into the overlooked parameter space of eccentric planets. Additionally, high-resolution spectroscopy has shown that the accreted material is more consistent with the composition of solar-system asteroids or the rocky planets than comets \citep{Zuckerman:2007tg, Klein2010}. As a result, we investigate the region near to the star interior and exterior to planetary orbit, as opposed to the distant planet and outer belt of material used in \citet{Bonsor2011}.  Furthermore, we simulate the entire region near the planet unlike the very detailed, single-MMR approach of \citet{Debes2012}. These simulations allow us to better understand the ability of a single planet to account for observed WD pollution rates, particularly as a function of planetary mass and eccentricity.

The layout of the paper is as follows: In Section 2 we review important features of the dynamics we expect to occur in the systems simulated. In Section 3 we discuss the setup and results of our initial simulation, and repeat that for a range of masses and eccentricities in Section 4 where we find that smaller and more eccentric planets are more efficient at delivering material to the host star. Section 5 details the theoretical effects of stellar evolution, and Section 6 presents the results of our WD simulations along with a comparison to the MS simulations, which show that stellar evolution can result in a significant population of newly unstable bodies. We finish with a discussion of results and comparison to other work in Section 7 and a conclusion in Section 8.

\section{System Dynamics}\label{sec:dyn}
The classical orbit of a single body in the gravitational field of a star will be a Keplerian ellipse, fixed in space. However, the orbit of a third, small body, such as an asteroid, will be affected by both the star and the planet and as a result be non-integrable. The orbit of such a body can be rapidly and dramatically changed upon close encounters with the planet, leading to ejection or collision with the star. In the case of bodies much less massive than the planet, the planetary orbit will remain unchanged. Beyond scattering, the orbits of small bodies can be perturbed by planets through both global secular effects and localized resonant effects. Through secular effects, orbits near an eccentric planet will slowly increase in eccentricity. This contribution is called `forced eccentricity': it is greatest for orbits near the planet and drops off with distance. As eccentricity increases so does the likelihood of scatterings, which further alters the orbit. 

Resonant effects occur at MMRs, and result from repeated encounters between bodies near the same location or locations over each orbit. These encounters can produce a rapid evolution of orbital elements such as eccentricity, or can act as a protection mechanism by preventing closer encounters between two bodies for long periods of time. MMRs are highly-localized and as such relatively small shifts in semi-major axis (SMA) can dramatically alter the stability of orbiting bodies, as will be shown in Section \ref{sec:msresults}. MMRs exist where the mean motions of two bodies form an integer ratio: 

\begin{equation}
\frac{n_2}{n_1} = \frac{p}{p+q}
\end{equation}
Here $n_1$ and $n_2$ are the mean motions of the inner and outer bodies, respectively, while $p$ and $q$ are integers. For orbits interior to the planet, the resonance is given by $p+q:p$ (so that $n_1$ is the mean motion of the planet), while for exterior orbits the resonance is $p:p+q$, with $n_2$ being the mean motion of the planet. The strength of the resonance is determined in part by the order of the resonance, $q$: smaller $q$ values generally correspond to stronger resonances. However, at larger planetary eccentricities higher-order MMRs are no longer negligible, which increases the number of trapping regions for small bodies. Particles located in resonances have been shown to become unstable at late times \citep{Wisdom:1982fj, Debes2012}, indicating that unstable MMRs will not be immediately cleared of bodies and can function as a source of material for the star even at late times. 

\subsection{The nominal chaotic zone}\label{sec:chaoticzone}
MMRs have a finite width that is determined by both the planet causing them, via its eccentricity and planet-to-star mass ratio, as well as the location and order of the resonance \citep{MurrayDermott}. As described in \citet{Chirikov:1979fj}, when resonances overlap in a system it results in stochastic motion and orbital instability. In the context of planetary systems, as $p$ increases and the mean motion of the orbit approaches that of the planet, first-order ($q=1$) MMRs are spaced more closely and eventually overlap. This region of overlap is known as the `chaotic zone' (CZ) and within it orbits are chaotic, frequently becoming unstable. \citet{Wisdom1980} showed that for a planet on a circular orbit, nearby bodies become unstable when the distance between them is

\begin{equation}\label{eq:chaoszone}
\varepsilon = \frac{a - a_p}{a_p}< 1.3 \mu^{2/7}
\end{equation}
where $a$ and $a_p$ are the SMAs of the nearby body and planet, respectively, and $\mu=$ M$_\mathrm{Pl}/$M$_*$ is the mass ratio between the planet and the central star. Numerical work by \citet{Duncan1989} showed the same mass dependence  with a slightly different coefficient:

\begin{equation}\label{eq:duncan}
\varepsilon_\mathrm{cz} = 1.5 \mu^{2/7}
\end{equation}
This equation indicates that larger planets produce larger CZs, for constant stellar mass. However, a corresponding equation for planets with moderate eccentricity does not exist: such a planet causes orbits of a third body to be non-integrable. Finally, this mass-ratio dependence also has important implications for the system as the star evolves off the MS, which will be covered in Section \ref{sec:evo}.

From the preceding paragraphs we can predict the general behavior of small bodies in the presence of a massive planet: Those closest to the planet will start between planetary pericentre and apocentre, and will rapidly be removed from the system by a collision or scattering in close encounter. Particles beyond the physical reach of the planet will be inside the CZ, and will likely go unstable as orbits become chaotic. Beyond the CZ, most particles should be stable with slight eccentricities, caused by secular effects. The exception is particles in MMRs, which may show anomalous behavior both in the CZ, where they may be more stable than their neighbors, and outside of the CZ, where they may be more unstable. Of those that are removed from the system, particles inside the orbit of the planet should have a higher likelihood of being accreted by the star, while those beyond the planet, being more weakly bound, should show a greater preference for ejection.  

\section{The first simulation}\label{sec:firstsim}

Due to the non-integrable nature of planetary systems with an eccentric planet and other bodies, we required numerical simulations to determine the orbital behavior of the small bodies.  We began by simulating a MS star and a planet of moderate eccentricity, for the purposes of comparing the results to our predictions from Section \ref{sec:dyn}. While we are primarily interested in the accretion rates around WDs, simulating a planet around both a MS and WD star allowed us to compare the dynamics of the system and understand the effect of stellar evolution.

\subsection{Setup}
Our first simulation was composed of a solar-mass star, a single 0.3 M$_\mathrm{Jup}$ planet orbiting with $a_p=4$ au and $e=0.2$, and 500 massless TPs distributed throughout the system. The stellar mass was selected due to the high frequency of solar-mass stars relative to more-massive stars, as well as for consistency with prior papers on the subject, specifically \citet{Bonsor2011} and \citet{Debes2012}. The planetary parameters were chosen to match theoretical predictions: exoplanets near to their host star will not survive post-MS evolution \citep{Rasio:1996lr}, but distant planets are unlikely to direct as much material to the star. To balance each factor the SMA was chosen to be 4 au, while the mass was selected to be in the middle of the observed eccentric exoplanet population, similar to $0.27$ M$_\mathrm{Jup}$ eccentric Saturn-analog OGLE-2006-BLG-109L c \citep{Gaudi:2008fv, Bennett:2010la}. 

The TPs were spaced 0.02 au apart from 0.06 au to 10 au,  allowing us to determine stability at a range of locations. While the planet was placed on an eccentric orbit, the TPs were placed on circular orbits at random mean longitudes. This decision was made for simplicity, and represents the case where secular effects have not had time pump the eccentricities of small bodies. Such a situation could occur in a young disc recently void of gas, or if the planet gained eccentricity impulsively (as would be the case in the model of \citet{DebesSigurdsson}). Non-zero initial eccentricities are discussed later in Section \ref{sec:QFcomp}. In addition, each TP was assigned a random inclination within 0.5$^\circ$ of the planet to avoid the artificial constraint of perfect coplanarity. The particles were assumed to be massless because of the minute mass estimated to be accreted by polluted WDs ($\sim5\times 10^{-7}$ M$_\mathrm{Jup}$) compared to our planet size, indicating a negligible effect by polluting material on the planet.

We also ran separate simulations of the strongest MMRs in greater detail: three interior to the planet at 3:1, 2:1, and 3:2; three exterior at 2:3, 1:2, and 1:3; and one co-orbital. We populated these resonances over a width of 0.2 au with a higher density of TPs,  spacing them 0.004 au apart, and assigned them the same orbital properties as above. Additionally, we simulated a second set in the same locations with the same spacing but with an initial eccentricity of 0.2 to match the planet. The initial longitude of pericentre of these remained random for consistency between all simulations. These non-circular TPs allowed us to determine the sensitivity of processes and results on initial eccentricity, particularly if it could lead to increased stability over the duration of the simulation. Additionally, by comparing these two very closely spaced populations we were able to test the dependence of the loss mechanism on the initial mean longitude, which was random for all particles.

During the simulations TPs were removed via one of three mechanisms: ejection from the system, occurring at 100 au; collision with the planet, which had a radius of 0.65 R$_\mathrm{Jup}$ given the assumed planetary density of 1.33 g cm$^{-3}$; or accretion by the central body, occurring when particles approached within 0.005 au. TPs that survived the $\sim$100 million year duration of the simulation were considered stable. The ejection radius was chosen to limit computation time, and it was found that increasing it to 1000 au had a negligible effect on the simulation results: the stability of particles was identical and the average lifetime in log space increased by less than five per cent. It should be noted that the majority of polluted WDs show cooling ages greater than the duration of our simulations, ranging from $10^{7.5}$ to $10^{10}$ years \citep{Debes2012}. Such time-spans were computationally too expensive to run for this work, so behavior late in the simulations ($10^6$--$10^8$ years) was used as a proxy. 

\subsection{Computation}\label{sec:comp}
We ran the simulation, as well as those in Sections \ref{sec:msresults} and \ref{sec:WDsims}, on the UCLA Institute for Digital Research and Education (IDRE) Hoffman2 cluster using the \textsc{Mercury} integrator package \citep{Chambers1999}. The package contains five N-body algorithms; we initially chose the hybrid symplectic integrator, which primarily uses a second-order mixed-variable symplectic algorithm and switches to a Bulirsch-Stoer (BS) algorithm upon close encounters, in our case 3 Hill radii. This combination has the advantage of shorter integration times than non-symplectic algorithms (e.g. BS) while still allowing close encounters, which are crucial for the scattering and accretion of TPs.  

To test the accuracy of the hybrid integrator we repeated the simulation using the non-symplectic BS integrator. We found that, while the former was faster and accurate in determining the stability of TPs in our simulations, the loss mechanism for a given particle was frequently inconsistent with the latter. The difference originates from the known issue that the hybrid integrator has difficulty with very large eccentricities \citep{Chambers1999}, as in the case of accreted particles. As a result, particles are  ejected when they should be accreted, as shown in Figure \ref{fig:comparealgs}. Given the importance of the accretion fraction on the stellar accretion rate, it was necessary for us to use the most accurate method available within the constraints of computation time. Therefore we used the hybrid simulation to determine the size of the unstable zone around the planet, which defined our eccentric chaotic zone (ECZ), and followed with a simulation of the ECZ using the BS integrator. We defined the edges of the ECZ as the region nearest to the planet having 19 out of 20 adjacent TPs stable (corresponding to a 0.4 au wide region with at least 95 per cent of small bodies stable), to ideally include all major instabilities near the planet. We integrated the MMRs using only the BS integrator, as they were narrow and unique in stability relative to the surrounding region. 

\subsection{Results}\label{sec:singleresults}
\begin{figure}
\begin{center}
\subfigure{
\includegraphics[width=.5\textwidth]{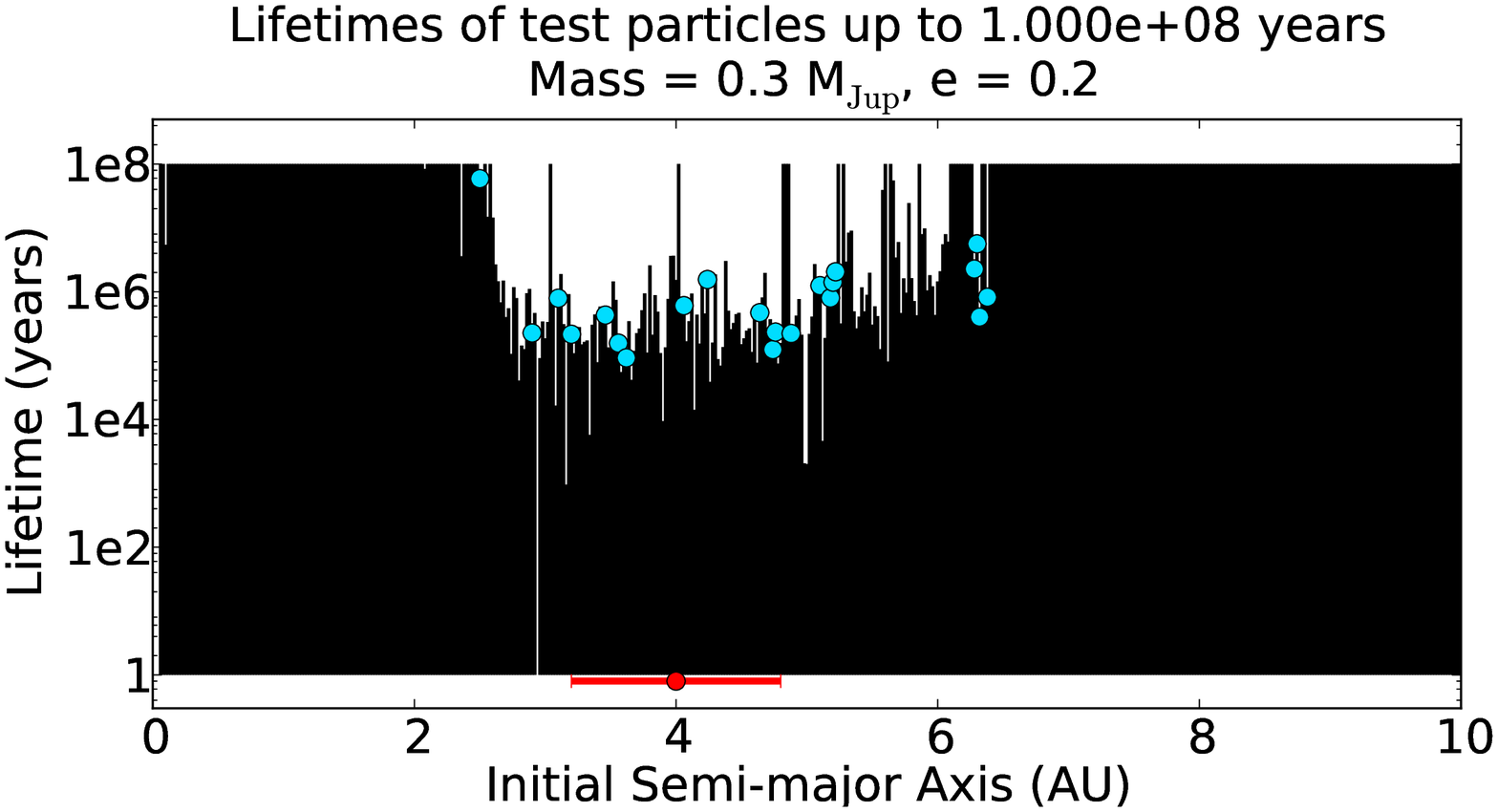}
\label{fig:comparealgs_a}}
\subfigure{
\includegraphics[width=.5\textwidth]{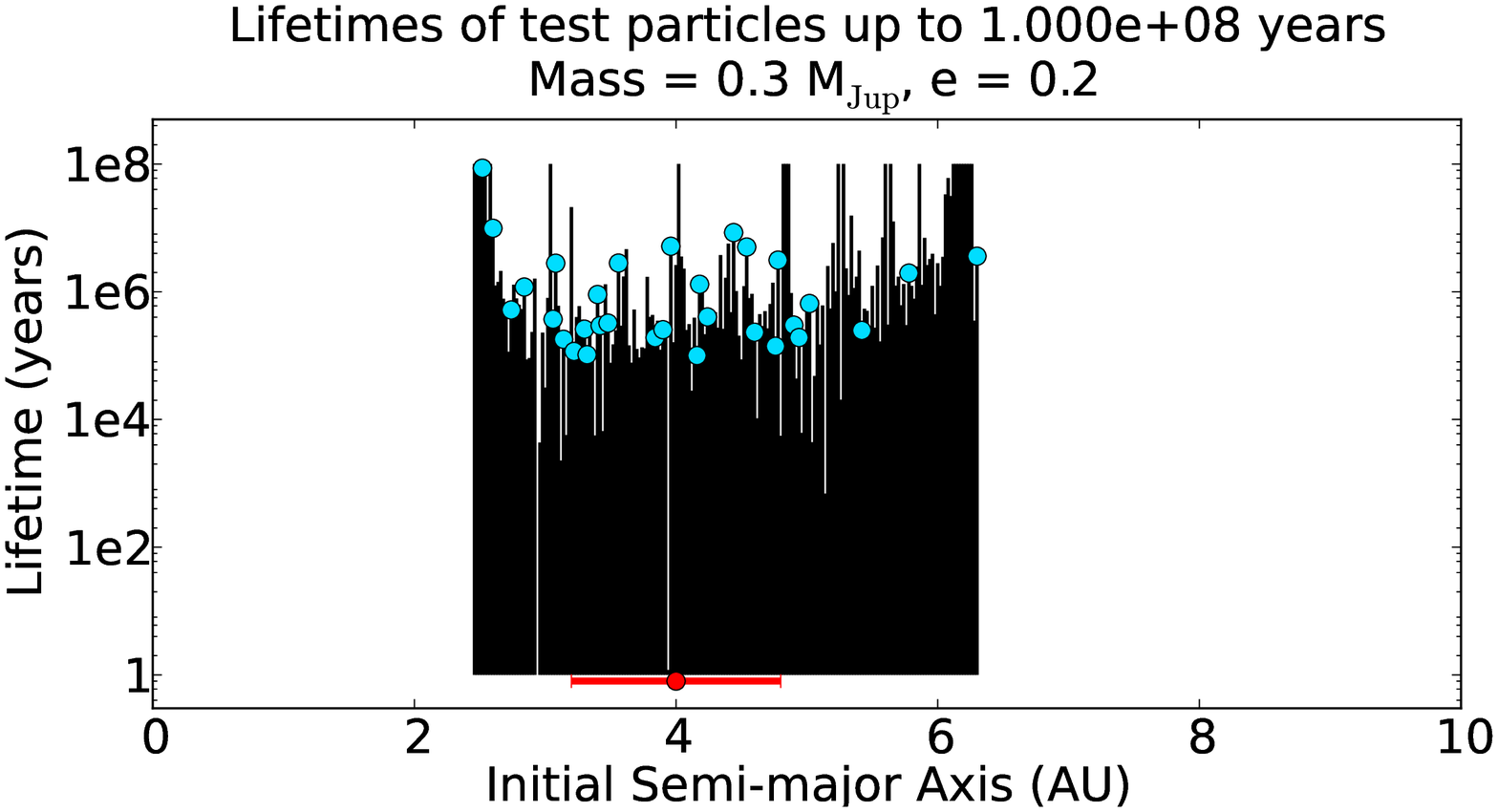}
\label{fig:comparealgs_b}}
\caption{Comparison of hybrid integrator (top) and BS integrator (bottom) results, showing particle lifetime as a function of initial SMA. Note the similarity in stability of TPs (black lines) around the planet (red point, error bars for pericentre and apocentre), but the difference in number of accreted particles (cyan points): 21 particles in the hybrid case and 31 in the BS case.  Simulations using the BS integrator were limited to regions with unstable particles to reduce computation time, as described in Section \ref{sec:comp}. \label{fig:comparealgs}}
\end{center}
\end{figure}

Upon completion, the simulation returned the orbital elements as a function of time, the loss mechanisms, and the lifetimes of each individual TP. As shown in Figure \ref{fig:comparealgs}, TPs that started near the planet rapidly went unstable, forming the ECZ between the range of $\sim2.4$ and $\sim6.3$ au.  Inside of this region some islands of stability existed, corresponding to low-order MMRs such as 2:1 and 3:4 at 3.05 and 4.85 au, respectively. The size of this ECZ was much larger than the predicted CZ for a planet on a circular orbit, which according to Equation \ref{eq:duncan} should have  extended from 3.4 to 4.6 au.
 
 \begin{figure} 
\begin{center}
\includegraphics*[width=0.5\textwidth]{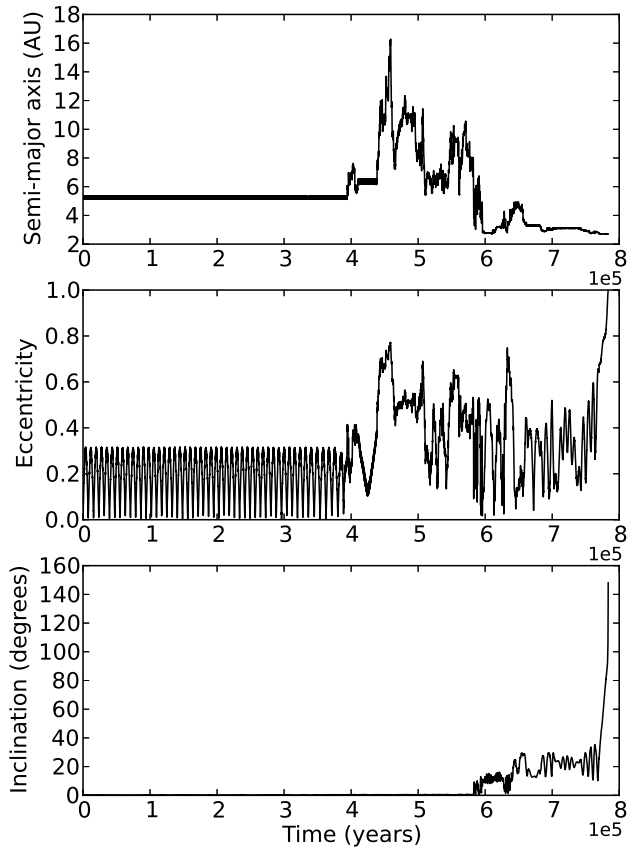}
\caption{Individual particle motion for an accreted TP starting at 5.3 au with $e=0$, showing periods of chaotic motion and apparent entrapment in MMRs. Also note the large increase in inclination before accretion occurred, which was common for accreted particles in the presence of the $e=0.2$ planet.  \label{fig:individual}}
\end{center}
\end{figure}

Individual particle motion depended on interactions with the planet and varied not only between TPs, but over the course of the simulation as well. One particle, representative of many others, started at 5.3 au and remained in the 2:3 MMR for nearly 400,000 years before scattering and undergoing chaotic motion (Figure \ref{fig:individual}). This particle appeared to be trapped in other MMRs during the remainder of the simulation, including the 1:2 resonance at 425,000 years and the 2:1 resonance at 725,000 years, shortly before accretion. Many unstable particles showed similar orbital motion that varied dramatically over the course of the simulation, spending some time apparently trapped in MMRs between periods of more chaotic motion, eventually being ejected or accreted as the eccentricity approached unity. Particles that collided with the planet also showed such motion, but failed to reach highly-eccentric orbits before being removed early in the simulation.

The final fates of particles were influenced by their starting locations: Particles that were accreted generally retained an apocentre near the planet, increasing in eccentricity through scattering until pericentre reached the surface of the star. Particles that were ejected often did so with a pericentre near the planet, as both the SMA and eccentricity increased. Because of this behavior, a larger fraction of unstable particles starting interior to the planet were accreted (24 per cent) than exterior (14 per cent). However, particles were lost via all three mechanisms in both starting regions due to chaotic motion.

\begin{figure} 
\begin{center}
\includegraphics*[width=0.5\textwidth]{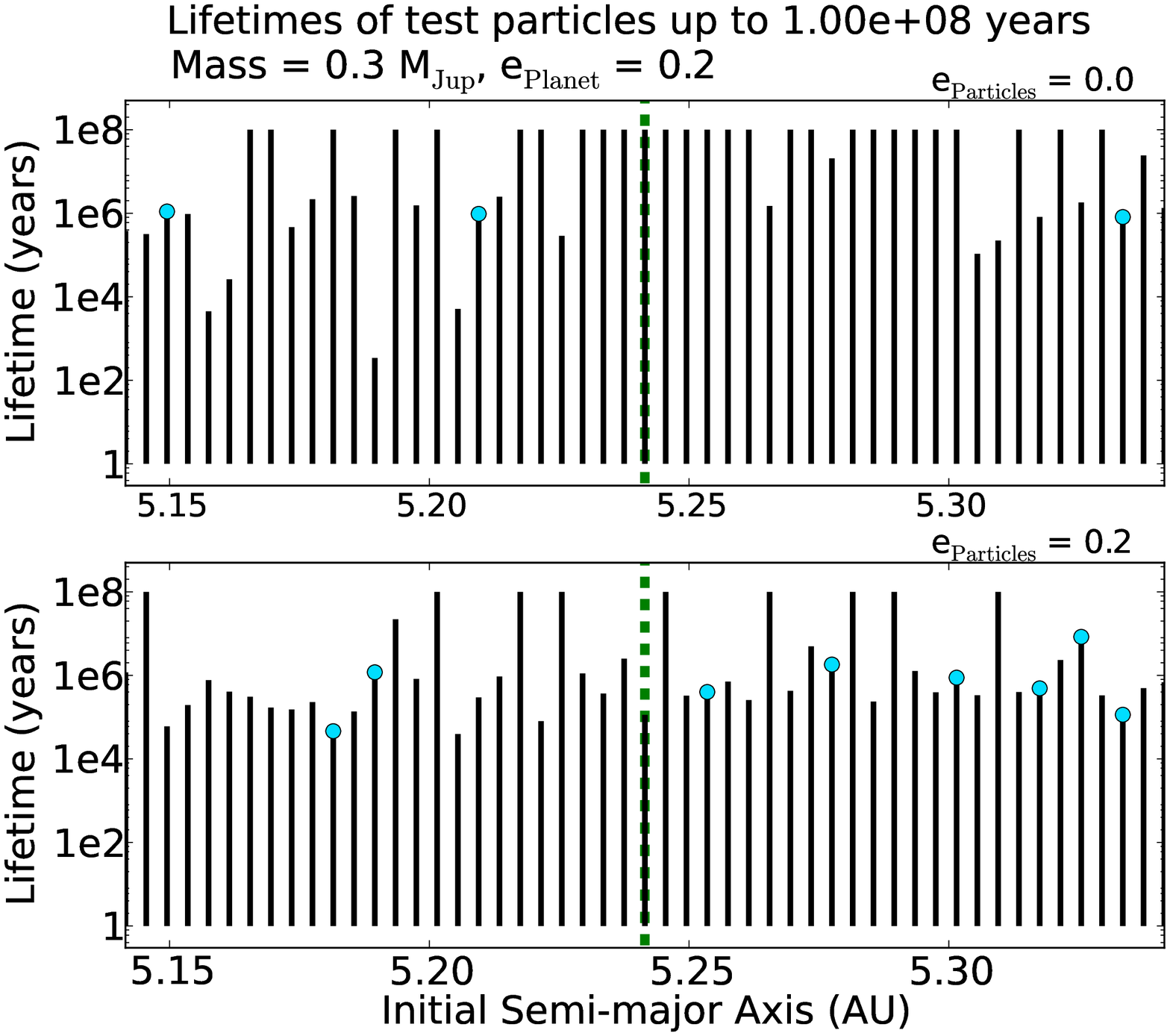}
\caption{Lifetimes of the particles around the 2:3 MMR (dashed line), initially circular orbits on top and initially eccentric orbits ($e=0.2$) on bottom. The circular particles are consistent with the results from the global simulations, while the eccentric particles were on average shorter-lived; this result supported our use of the initially-circular TPs in the other simulations. \label{fig:res23}}
\end{center}
\end{figure}

The higher-density MMR simulations were analysed separately and showed consistency with the above results, including behaving as islands: particles starting very near to some low-order MMRs, such as the 2:3 and 3:2 resonances, were much more stable than the surrounding particles (Figure \ref{fig:res23}, top). Additionally, the 2:1 resonance was found to be a major source of accreted particles at late times. Within 0.1 au of the resonance, 15 out of 16 unstable particles were accreted and 11 of those survived more than 1 million years. The $e=0.2$ populations in each MMR were also stable in some of the same locations, but in general had shorter lifetimes. This result was evidence that such particles would be more rapidly removed from a planetary system, and TPs that were started on circular orbits more accurately represented the material around a star that would exist at later times. Since our primary interest was the behavior of TPs at late times, this result reinforced our decision to use the particles on circular orbits for the large scale simulations.

\section{Results for range of eccentricities and masses}\label{sec:msresults}
To investigate the effect of planetary properties on the stability of surrounding bodies, we repeated the previous simulation for each combination of four planetary masses and five orbital eccentricities. We chose the mass values 0.03, 0.3, 1.0, and 4.0 M$_\mathrm{Jup}$ and the eccentricities 0.02, 0.2, 0.4, 0.6, 0.8 to probe the wide range in orbits of known exoplanets. Aside from eccentricity and mass, all planets started with the same initial conditions in all simulations, such as mean anomaly and argument of pericentre. The SMA remained 4 au for all simulations as well, as probing a third dimension of parameter space would have been too computationally expensive. 

\subsection{Mass Effects}\label{sec:MSmass}

\begin{figure*}
\begin{center}
\includegraphics*[width=1\textwidth]{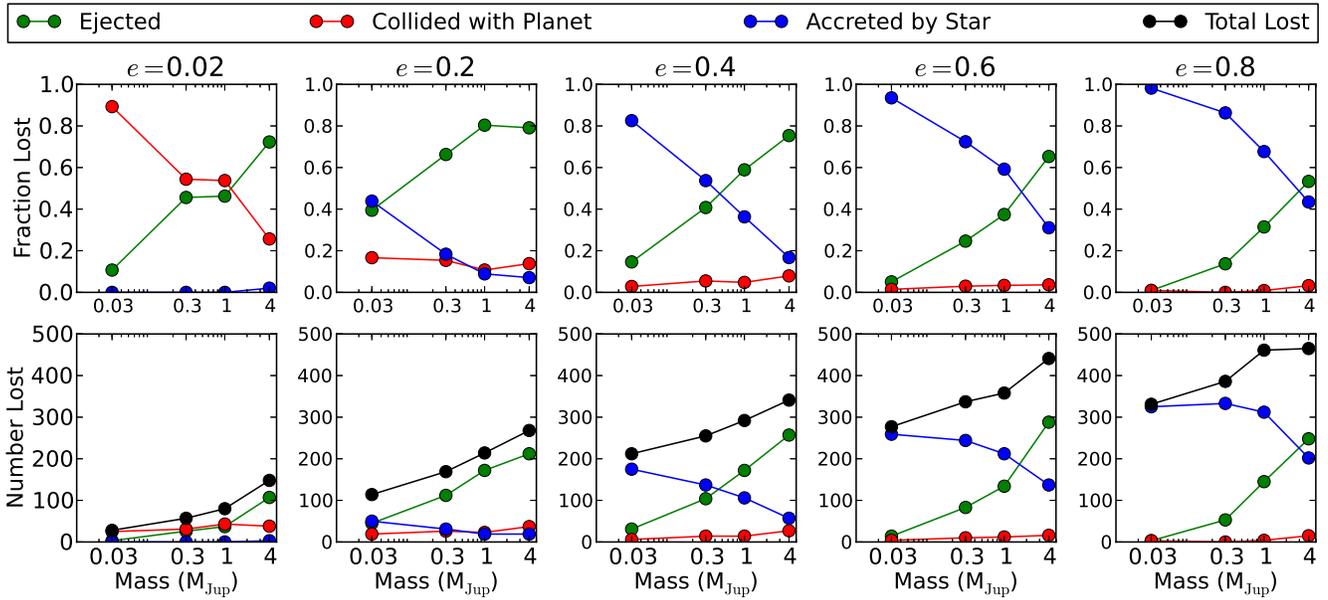}
\caption{(top) Fraction of unstable TPs lost by each mechanism as a function of planetary mass, $\mathrm{N}_\mathrm{mechanism}/\mathrm{N}_\mathrm{Lost}$, for each planetary eccentricity over $10^8$ years. In all cases with $e\ge0.2$ the accretion fraction is a monotonically decreasing function of the mass. (bottom) Total number of particles lost by each mechanism. Despite increasing the amounts of unstable particles, increasing planetary mass resulted in fewer accreted particles. \label{fig:msmasslost}}
\end{center}
\end{figure*}

\begin{figure*}
\begin{center}
\includegraphics*[width=1\textwidth]{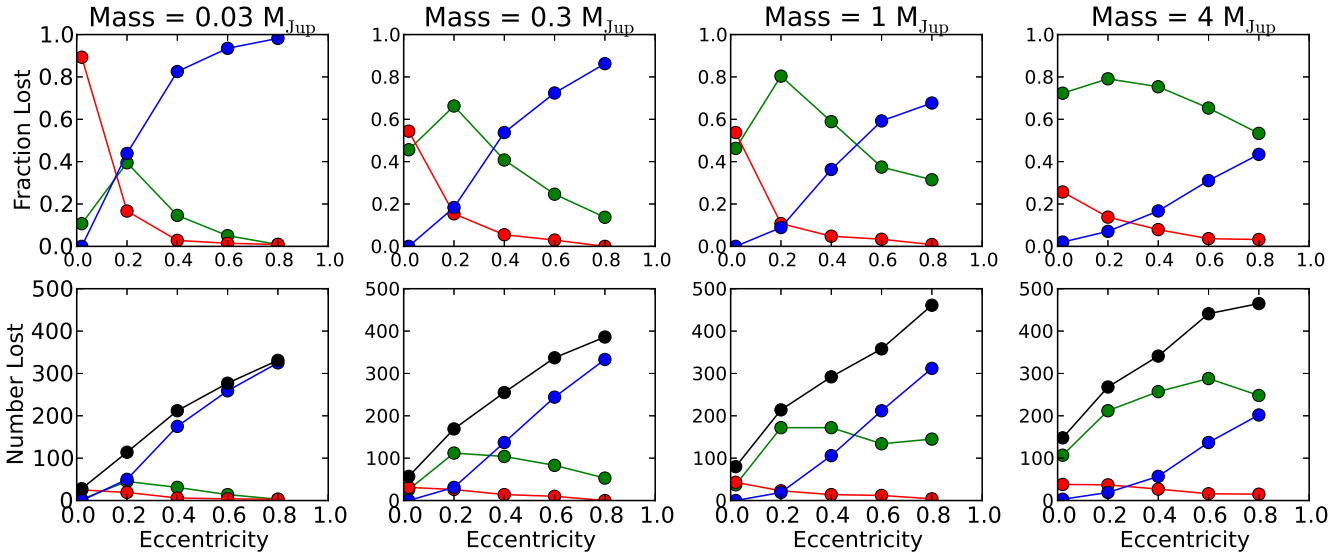}
\caption{(top) Fraction of unstable TPs lost by each mechanism as a function of planetary eccentricity, for each planetary mass over $10^8$ years. The dramatic and monotonic increase in accretion fraction with eccentricity is clear over all masses. (bottom) Total number of particles lost by each mechanism. Due to the growth of the ECZ, the increase with eccentricity is even greater. \label{fig:mseccfraclost}}
\end{center}
\end{figure*}

Our simulations showed that the total number of particles accreted by the star decreased with increasing planetary mass for all but the most eccentric planets (Figure \ref{fig:msmasslost}, bottom, blue lines). Two factors contributed to this weak dependence: the size of the unstable region and the fraction of unstable particles accreted by the star. We found that while a larger planetary mass corresponded to a larger ECZ  and a higher total number of unstable particles (black lines), it dramatically reduced the fraction accreted (Figure \ref{fig:msmasslost}, top, blue lines). This reduced fraction more than offset the increased ECZ size, reducing the total number of particles accreted. The physical cause of this relationship is the strength of gravitational interactions: while the $\ge 1$ M$_\mathrm{Jup}$ planets rapidly ejected most particles, smaller planets repeatedly interacted more weakly with the TPs and allowed them to slowly diffuse inward and be accreted by the star. 

 \begin{figure} 
\begin{center}
\includegraphics*[width=0.5\textwidth]{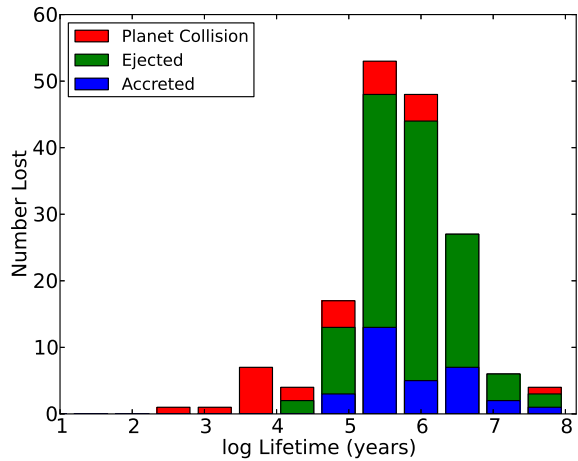}\\
\includegraphics*[width=0.5\textwidth]{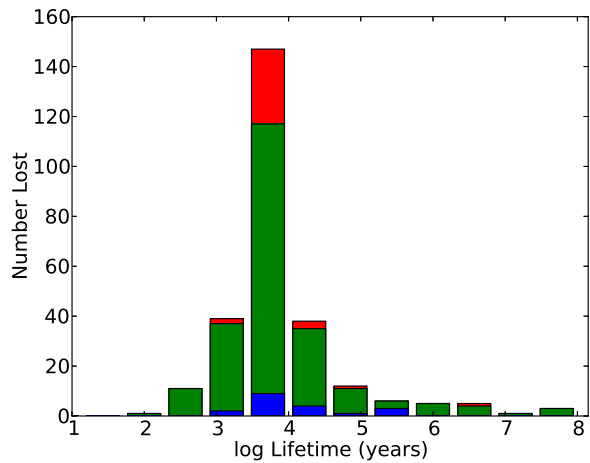}
\caption{A comparison of the lifetimes for TPs lost in a system with a lower mass planet (0.3 M$_\mathrm{Jup}$, top) and a massive planet (4 M$_\mathrm{Jup}$, bottom), both with planetary eccentricity $e=0.2$. The smaller planet resulted in longer lifetimes for unstable TPs as well as a larger fraction lost via accretion (blue bars). \label{fig:histlifetimes}}
\end{center}
\end{figure}

In addition to changing the fraction of TPs accreted and ejected, planetary mass also affected the time-scale for instability to set in. Physically, more-massive planets are more capable of removing small bodies after a single scattering event, while smaller planets depend on the cumulative effect of multiple scatterings. As a result, the TPs in simulations with massive planets had shorter lifetimes than those in simulations with small planets. Figure \ref{fig:histlifetimes} illustrates this effect with the lifetimes of TPs perturbed by the most massive planet (4 M$_\mathrm{Jup}$) and a lower mass planet (0.3 M$_\mathrm{Jup}$); Section \ref{sec:latetimeaccr} examines the characteristic lifetimes of particles in greater detail. 

\subsection{Eccentricity Effects}\label{sec:ecc}
The planetary eccentricity had an equally powerful effect on the TPs. As it increased the pericentre and apocentre of the planet shrank and grew, respectively, which widened the region around the star that the planet probed and expanded the ECZ. As a result, the planet caused a larger number of TPs to be destabilized and accreted (Figure \ref{fig:mseccfraclost}, bottom, black lines). We also found that the eccentricity affected the fraction of TPs accreted, in fact more strongly than the planetary mass (Figure \ref{fig:mseccfraclost}, top, blue lines). Physically, eccentric planets drive the forced eccentricity of particles up to higher values, while simultaneously having more and wider MMRs that can increase the eccentricity of the particles in them. Higher eccentricity causes a higher accretion probability. Finally, increasing planetary eccentricity resulted in the stellar accretion rate peaking earlier, as characterized by the mean accretion time in log space. We discuss this effect  in greater detail and with respect to WD accretion in Section \ref{sec:WDsims}. 

In the nearly-circular ($e=0.02$) runs, planet-TP collisions dominated the loss mechanism, particularly at low mass. We were not hugely surprised by this behavior, as few TPs had eccentricities pumped to values large enough to eject or accrete. In these simulations most particles maintained a nearly constant Tisserand parameter, defined as 
\begin{equation}
T=\frac{1}{a/a_p} + 2\sqrt{\frac{a}{a_p}(1-e^2)}\cos I
\end{equation}
Here $a$ and $a_p$ are the SMAs of the particle and planet, respectively, while $e$ and $I$ are the particle eccentricity and inclination relative to the planetary orbit \citep{MurrayDermott}. For small inclinations, an increase in eccentricity requires the SMA to increase as well. The Tisserand parameter determines the region of parameter space a particle can explore in the presence of a planet with zero eccentricity, including a minimum pericentre. 

As described in \citet{Bonsor:2012bh}, for an accretion distance of 0.005 au and an ejection distance of 100 au only particles with $T <  2.1$ or $T<2.85$ can be accreted or ejected, respectively. With the initial conditions $e=0$ and $\cos I\approx1$, all TPs started out with $T>3$ and were incapable of close approach with the central star. The small eccentricity of 0.02 was enough, however, for particles to deviate slightly from a constant Tisserand parameter and be ejected. Ejections were most common in simulations with the more-massive planets, which were capable of ejecting particles even when on circular orbits \citep{Bonsor:2012bh}. For planetary eccentricities larger than 0.02, the Tisserand parameter was not a constant of motion (due to being derived from the restricted 3-body problem, where the planet is on a circular orbit). As a result, particle eccentricity was  frequently increased with no corresponding growth in SMA.

\subsection{ECZ width}\label{sec:ECZ}
\begin{figure}
\begin{center}
\includegraphics*[width=0.5\textwidth]{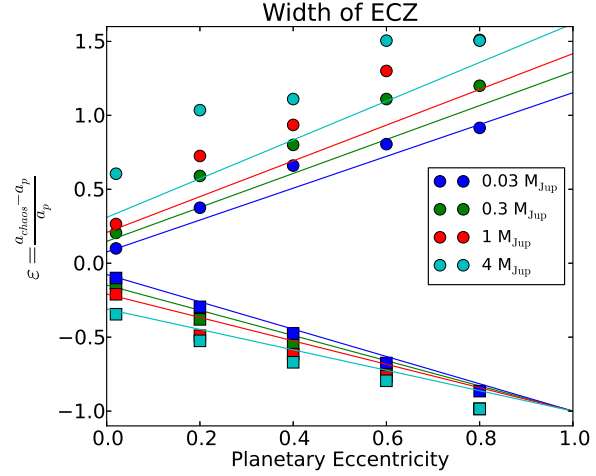}
\caption{Scaled distance to the interior (squares) and exterior (circles) edges of the ECZ, as a function of eccentricity and planetary mass. The lines represent our interior and exterior model fits: $\varepsilon_\mathrm{in} = e_\mathrm{p} (1 - \varepsilon_\mathrm{cz}) + \varepsilon_\mathrm{cz}$ and $\varepsilon_\mathrm{out} = e_\mathrm{p} (1 + \varepsilon_\mathrm{cz}) + \varepsilon_\mathrm{cz}$. \label{fig:ECZwidth}}
\end{center}
\end{figure}

In the case of a planet on a circular orbit, the interior and exterior edges of the CZ should be equal and scale simply with $\mu^{2/7}$ as described in Section \ref{sec:chaoticzone}. In our simulations we found that both mass and eccentricity served to increase the size of the ECZ, as shown by the points in Figure \ref{fig:ECZwidth}.  Additionally, we found that the influence of planetary eccentricity served to produce interior and exterior edges at markedly different distances from the planet. To fit the eccentricity-ECZ effect we needed a model with three properties: showed edge asymmetry; increased with planetary eccentricity; and reduced to $\varepsilon_\mathrm{chaos}=1.5\mu^{2/7}$ for $e_\mathrm{p}=0$.

From a physical standpoint, a particle can be removed from the system when it crosses the path of the planet. Particles within the CZ in the zero-eccentricity case are simply those that undergo chaotic motion, and can therefore enter the path of the planet. For a given planet eccentricity and particle eccentricity, orbital crossing occurs at 
\begin{align}
a_\mathrm{in} &= a_\mathrm{p} (1- e_\mathrm{p}) / (1 + e_\mathrm{in}') \\ 
a_\mathrm{out} &= a_\mathrm{p} (1+ e_\mathrm{p}) / (1 + e_\mathrm{out}')
\end{align} 
where $a_\mathrm{in}$ ($a_\mathrm{out}$) is the inner (outer) edge of the ECZ, and $e_\mathrm{in}'$ ($e_\mathrm{out}'$) is the characteristic eccentricity leading to instability for the inner (outer) region. In this case,
\begin{align}
\varepsilon_\mathrm{in} &= \frac{e_\mathrm{p} + e_\mathrm{in}'}{1+ e_\mathrm{in}'}, &
\varepsilon_\mathrm{out} &= \frac{e_\mathrm{p} + e_\mathrm{out}'}{1 - e_\mathrm{out}'}
\end{align}

To determine $e_\mathrm{out}'$ and $e_\mathrm{in}'$, both edges were assumed to obey the condition $\varepsilon(e_\mathrm{p} = 0)=\varepsilon_\mathrm{cz} = 1.5\mu^{2/7}$, the size of the CZ as determined by \citet{Duncan1989} for a planet on a circular orbit. Substituting for $e_\mathrm{out}'$ and $e_\mathrm{in}'$, the edges of the ECZ are defined by
\begin{align}\label{eq:ECZ}
\varepsilon_\mathrm{in} &= e_\mathrm{p} (1 - \varepsilon_\mathrm{cz}) + \varepsilon_\mathrm{cz}, &
\varepsilon_\mathrm{out} &= e_\mathrm{p} (1 + \varepsilon_\mathrm{cz}) + \varepsilon_\mathrm{cz}
\end{align}
as shown by the lines in Figure \ref{fig:ECZwidth}. While this function does not match our results perfectly, it does exhibit the asymmetry and expansion with eccentricity found in our simulation results, and does so with physical motivation. Finally, this model does not explicitly include the increase in resonance width with eccentricity and resulting change in resonance overlap, so it is unsurprising to see deviations from the fit.

\subsection{Mean motion resonances}\label{sec:MMRres}
\begin{figure}
\begin{center}
\includegraphics[width=.5\textwidth]{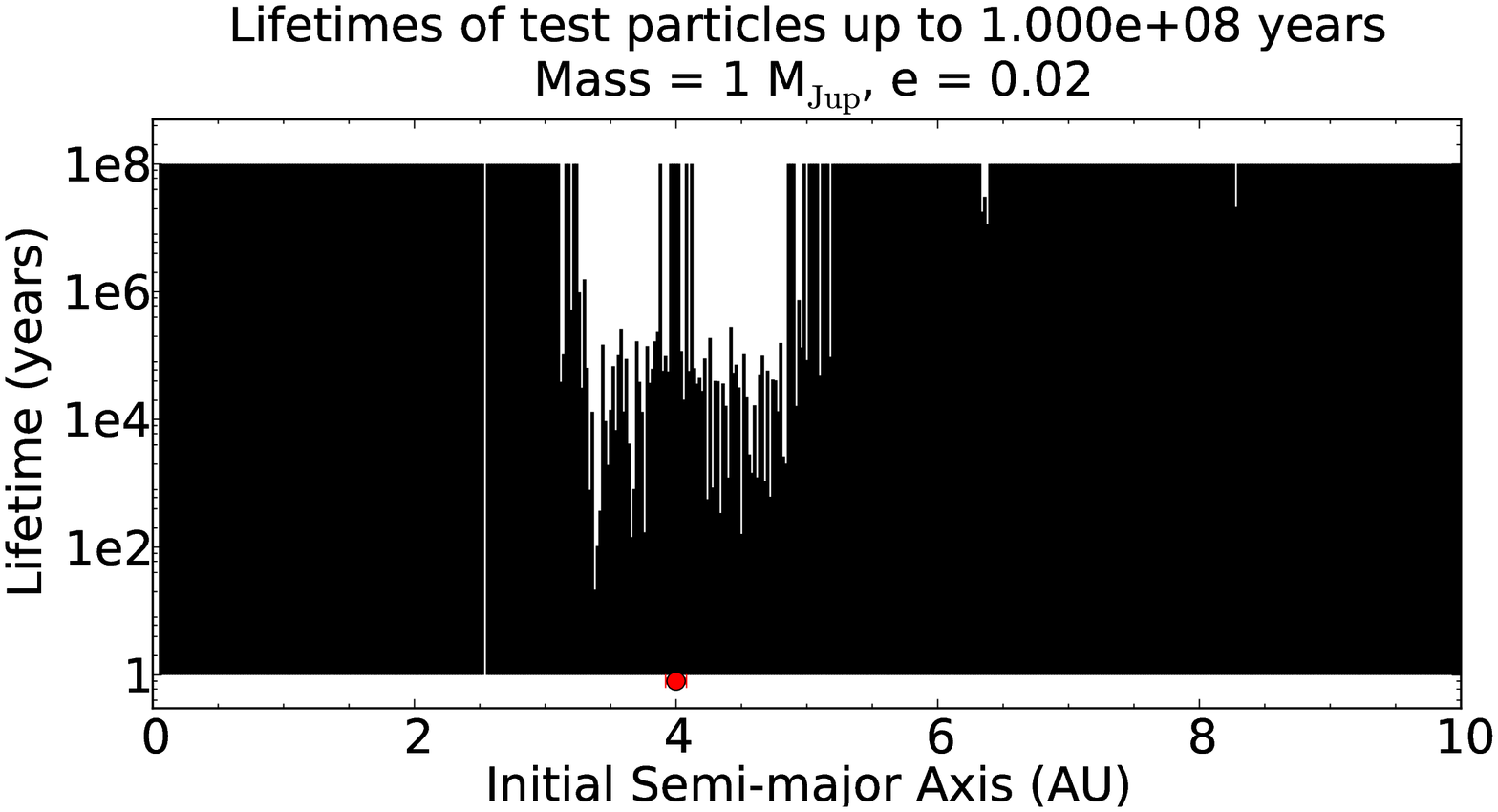}\\
\includegraphics[width=.5\textwidth]{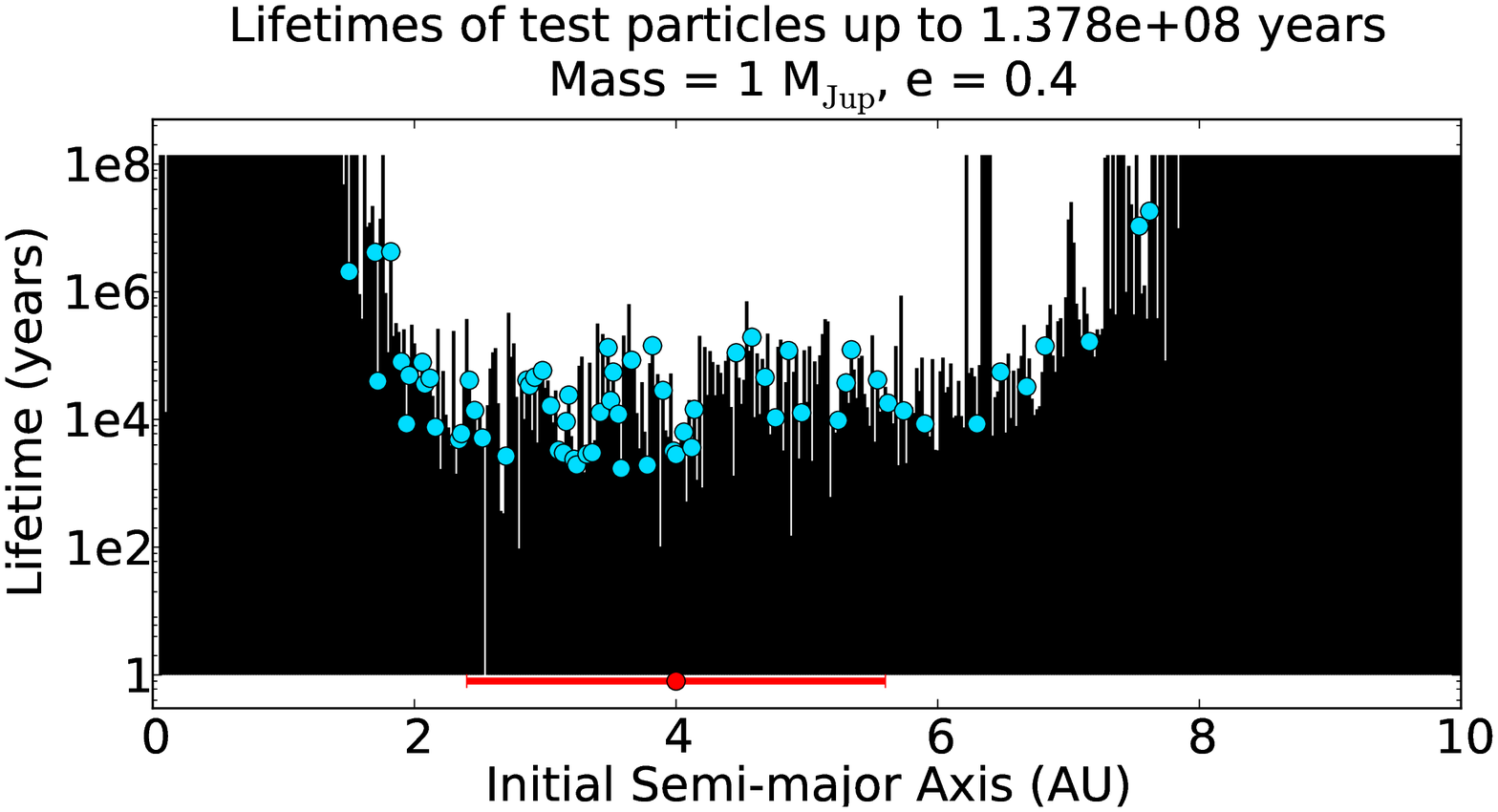}
\caption{Lifetimes of TPs for a 1 M$_\mathrm{Jup}$ planet for planetary eccentricity $e=0.02$ (top) and $e=0.4$ (bottom). Note the change in the stability of the 1:2 MMR at 6.35 au: in the low eccentricity case particles in the MMR are short-lived and unstable, while in the case of high eccentricity planet the MMR acts as a safe haven in a region of instability.\label{fig:MMRchange}}
\end{center}
\end{figure}

As described in Section \ref{sec:singleresults}, though many TPs that started in the vicinity of an MMR behaved similarly to the adjacent regions, some were stable inside of the ECZ (islands) and others were unstable outside of the ECZ (holes). Many of these holes became islands as the eccentricity of the planet increased and the MMR went from outside the ECZ to inside of it. In the case of 1 M$_\mathrm{Jup}$, the 1:2 resonance at 6.4 au changed from nearly the only unstable location outside of the ECZ at $e=0.02$ to the only region of instability inside the ECZ at $e=0.4$ (Figure \ref{fig:MMRchange}). As a result, these holes were more rare in the higher-eccentricity simulations, due to the fact that the nearly all first-order MMRs were contained within the ECZ. Even so, they played an important role in delivering material to the star for some planets. In the case of $e=0.2$ and the largest masses, 4 M$_\mathrm{Jup}$ and 1 M$_\mathrm{Jup}$, the 3:1 resonance had a much higher fraction of TPs accreted by the star. In the 4 M$_\mathrm{Jup}$ case, 56 per cent of unstable TPs near the MMR were accreted, significantly larger than the accreted fraction of all unstable particle interior to the  planet, 13 per cent.

We also found that the 1:1 resonance frequently retained TPs through the end of the simulation, for planets of eccentricity of 0.02 and 0.2. Most of these particles were on stable Trojan orbits. The mass of the planet determined the limiting eccentricity for Trojans. While the smallest planet (0.03 M$_\mathrm{Jup}$) had Trojans when the eccentricity was up to and including 0.4, the planets with mass 0.3 M$_\mathrm{Jup}$ and 1 M$_\mathrm{Jup}$ only saw Trojans for $e=0.2$ and 0.02. The most massive planet, 4 M$_\mathrm{Jup}$, only had Trojans at the lowest eccentricity.

Finally, we again ran simulations with TP eccentricity matching planetary eccentricity. These simulations, while generally less stable than those with circular TPs, exhibited the same dependencies on planetary eccentricity and mass as those with TPs on circular orbits in all cases. We therefore believe these relationships do not depend strongly on particle eccentricity, though they may depend on particle inclination or longitude of pericentre.  

\subsection{Comparison with previous work}\label{sec:QFcomp}
Other research has also found similar trends with respect to planetary mass, such as \citet{Bonsor2011}. In both cases, larger planets produced shorter TP lifetimes, higher ejection fractions, and reduced fractions either in the inner system (for \citet{Bonsor2011}) or accreted by the star (our result). While the total belt size differed in each case, we obtained consistent results for the total number of particles accreted as a function of planetary mass: where \citet{Bonsor2011} saw a weak decrease in the total mass scattered to the inner system as the mass of the planet increased, we saw the analogous result of fewer TPs accreted as the planet mass increased (at non-zero eccentricities).

Conversely, our results deviate from those of \citet{QuillenFaber2006}, who found that the size of the CZ is independent of planetary eccentricity up to values of 0.3 regardless of planetary mass. By running our simulations with the TP initial conditions changed to those of the prior authors (coplanar particles given the predicted forced eccentricity and planetary longitude of pericentre at their initial SMA), we matched their results and thus identified the different starting conditions as the source of discrepancy. While our TPs began with random longitude of pericentre, \citet{QuillenFaber2006} used the expected forced eccentricity at that location along with the same longitude of pericentre as the planet. Given that our simulations behaved similar for TPs with both planetary eccentricity and zero eccentricity, the difference is likely the longitude of pericentre in conjunction with the forced eccentricity, which can prevent close encounters. Additionally, coplanarity prevents particles from reaching inclined orbits, which often occurred for the unstable particles in our simulations.

\begin{figure}
\begin{center}
\includegraphics[width=.5\textwidth]{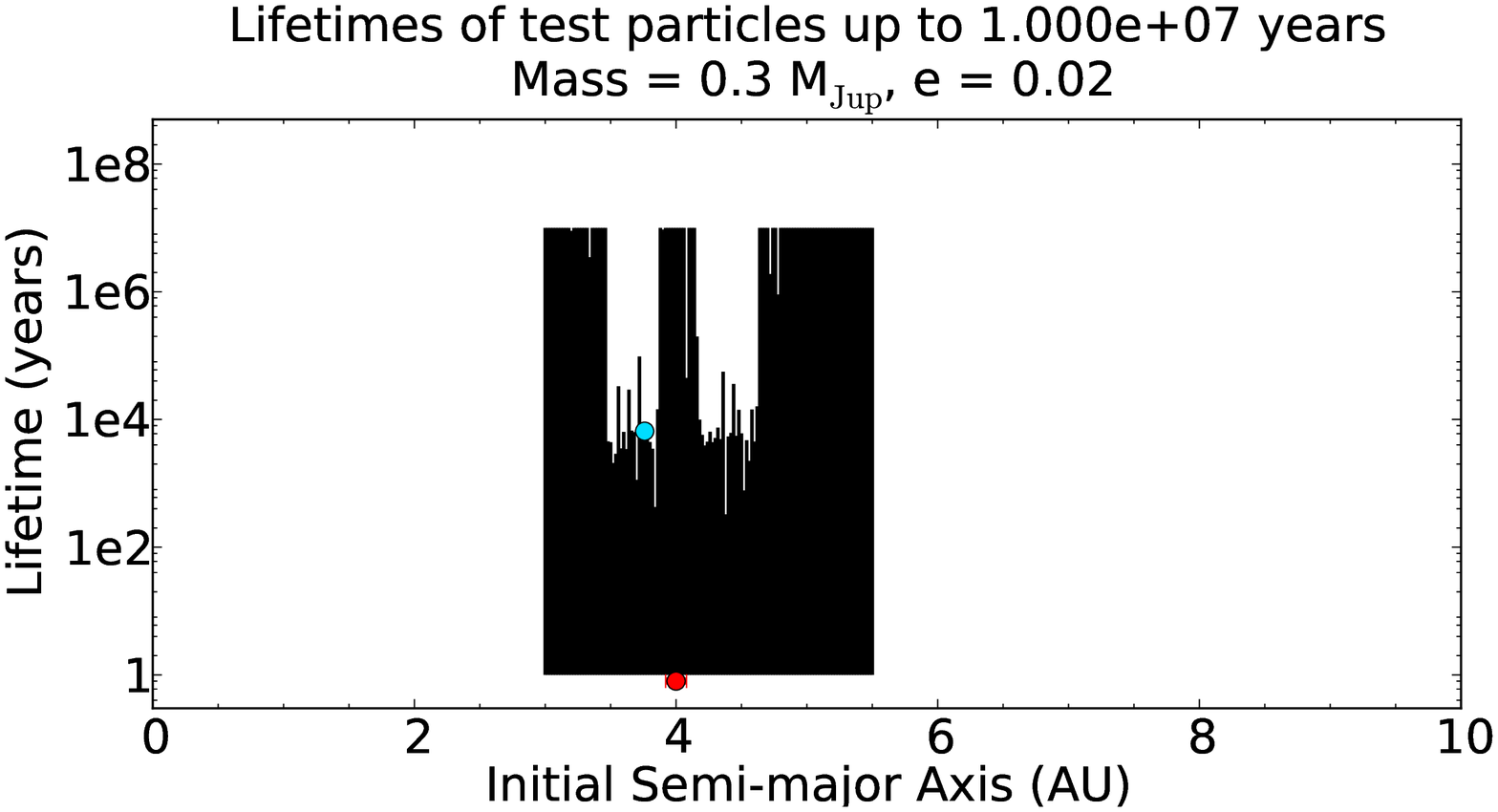}\\
\includegraphics[width=.5\textwidth]{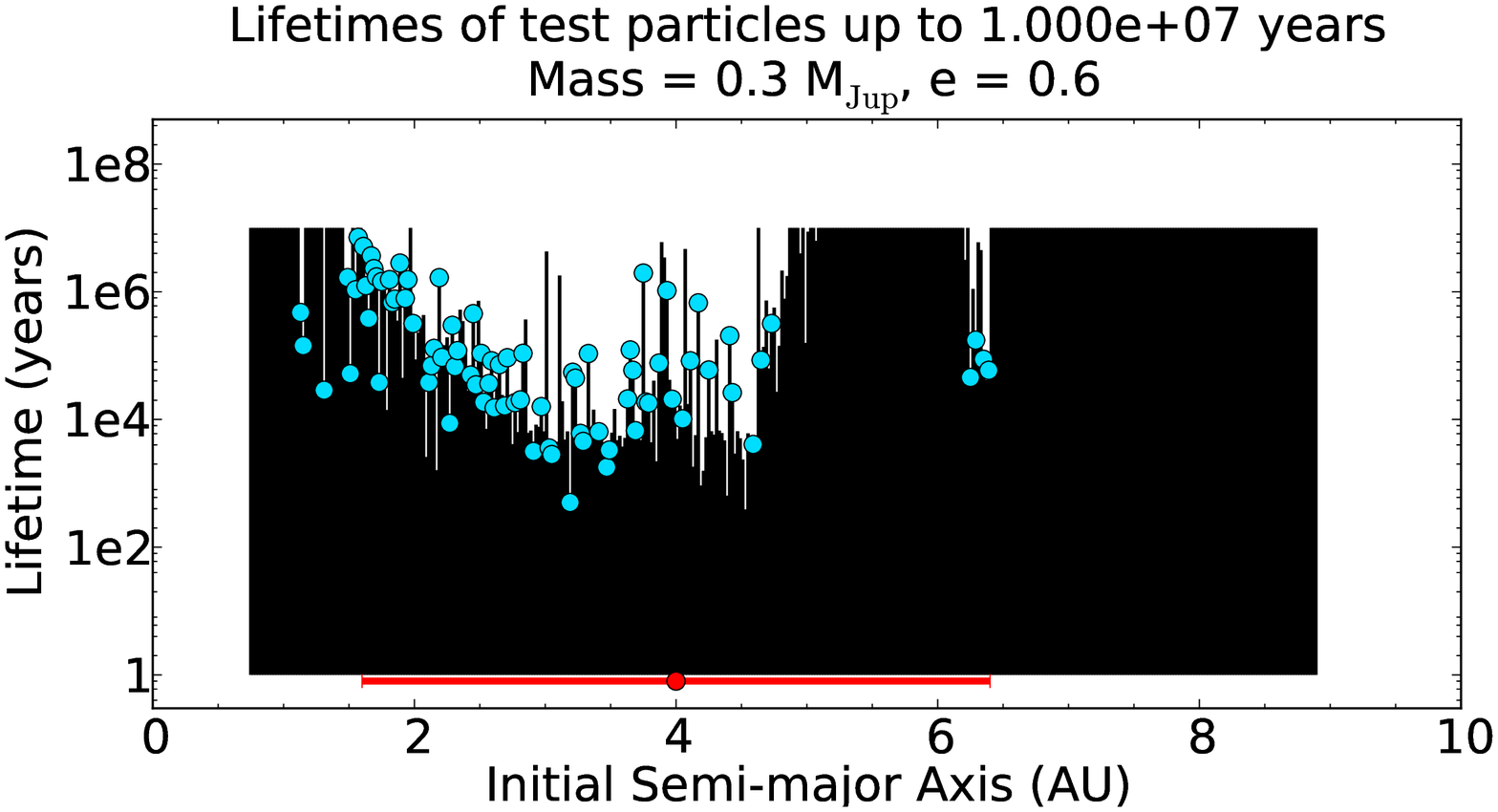}
\caption{Particle lifetimes in the presence of a 0.3 M$_\mathrm{Jup}$ planet with eccentricity $e=0.02$ (top) and $e=0.6$ (bottom). In these simulations the TPs began on orbits with the forced eccentricity at their starting location and the same argument of pericentre as the planet. In this case, the size of the ECZ changes, but the expansion is limited almost entirely to the region interior the planet. \label{fig:QFplots}}
\end{center}
\end{figure}

We also found that even with the changed initial conditions, planetary eccentricity larger than 0.2 still increased the size of the ECZ, albeit in a manner different from Figure \ref{fig:ECZwidth}. While the external edge remained close to the value expected from Equation \ref{eq:duncan}, the internal edge changed dramatically, resulting in a much larger ECZ (Figure \ref{fig:QFplots}). This effect was only noticeable in the larger eccentricities, which were not investigated in \citet{QuillenFaber2006}. Additionally, the effect of eccentricity on the number and fraction of accreted particles, as in Figure \ref{fig:mseccfraclost}, remained. The effect of planetary mass on accretion fraction remained as well, but was markedly weaker. As a result, the number of accreted particles was affected more strongly by the wider ECZ for massive planets, and increased with larger planetary mass. These simulations, along with the results described in Section \ref{sec:ECZ}, appear to indicate that the initial conditions of the particles play an important role in the some aspects of the system (such as size of ECZ), while other aspects are less affected (like the eccentricity-accretion relation). Therefore the importance of planetary eccentricity on WD accretion cannot be ignored.

\section{Stellar Evolution}\label{sec:evo}
Until this point we have examined the stability of TPs in planetary systems around only an unevolved MS star. However, to estimate the WD accretion rate we also need to account for the reaction of the planetary system to the mass loss that occurs during post-MS evolution. Research shows that stellar evolution can change the stability of some orbits in the system due to the large amount of mass lost during the asymptotic giant branch and to a lesser extent the red giant branch \citep{Mustill:2013fj, Veras:2013vn}. From a purely gravitational standpoint, the ECZ dependence on the planet-to-star mass ratio (Equation \ref{eq:duncan}) indicates that the ECZ will widen, due to the decrease in stellar mass. Additionally, the mass dependence of the MMR widths causes them to grow as well, as was shown in \citet{Debes2012}. 

The final mass of a WD can be estimated according to the equation for the empirical initial-final mass relation from \citet{Wood1992}:

\begin{equation}\label{eq:massloss}
\mathrm{M}_\mathrm{WD}=0.49\exp[0.095\mathrm{M}_\mathrm{MS}]
\end{equation} 
Here M$_\mathrm{MS}$ and M$_\mathrm{WD}$ are the initial (MS) mass and final (WD) mass, respectively, of the star in solar masses. For stars near a solar-mass star this equation gives a WD mass of 0.539 M$_\odot$, which is consistent with more recent work on the relation including \citet{Weidemann:2000wd} and \citet{Kalirai:2008eu}. 

Given adiabatic mass loss, conservation of angular momentum dictates the SMAs of orbiting planets will increase by an amount determined by the initial-to-final mass ratio \citep{DebesSigurdsson}:

\begin{equation}\label{eq:aexpand}
a_f=a_0\left(\frac{\mathrm{M}_\mathrm{MS}}{\mathrm{M}_\mathrm{WD}}\right)
\end{equation} 
For a body around a solar-mass star the orbit will expand by a factor of 1.86. A planet with an SMA of 4 au will then reach a new orbit of 7.42 au. Since this orbital expansion affects all bodies in the system equally (ignoring non-gravitational effects), the period ratios between them will remain the same and bodies already in MMR with a planet will remain so. However, the larger planet-to-star mass ratio results in larger widths for MMRs, which can result in new objects entering resonance. The change in central mass also affects the size of the ECZ. From Equation \ref{eq:duncan} we can predict the magnitude of this increase in the case of zero eccentricity:
\begin{equation}\label{eq:eczexp}
\varepsilon_\mathrm{WD}/\varepsilon_\mathrm{MS} = 
(\mu_\mathrm{WD}/\mu_\mathrm{MS})^{2/7} = 
(\mathrm{M}_\mathrm{MS}/\mathrm{M}_\mathrm{WD})^{2/7}
\end{equation}
For our star the mass loss should produce a widening of 19 per cent. However, we saw in Section \ref{sec:ECZ} that eccentric planets do not follow this relation closely. Assuming the edges of the ECZ are actually defined by Equation \ref{eq:ECZ}, which more closely matches our MS simulations, the relative growth would drop with increasing planetary eccentricity. As the eccentricity increases, the contribution from the classic CZ ($\varepsilon_\mathrm{cz}$) to the size of the ECZ decreases and the orbital excursions of the planet dominate. The ratio in that case is given by 
\begin{equation}\label{eq:eczexp2}
\varepsilon_\mathrm{WD}/\varepsilon_\mathrm{MS} = \frac{e_\mathrm{p}(1\pm 1.5\mu_\mathrm{WD}^{2/7}) + 1.5\mu_\mathrm{WD}^{2/7}}{e_\mathrm{p}(1\pm 1.5\mu_\mathrm{MS}^{2/7}) + 1.5\mu_\mathrm{WD}^{2/7}}
\end{equation}
As a result, more-eccentric planets with ECZs following this relation will have a relatively smaller amount of expansion. However, the physical growth of the unstable region ($\varepsilon_\mathrm{WD}-\varepsilon_\mathrm{MS}$) will increase with planetary eccentricity for the exterior edge, and the greater fraction of TPs accreted in the presence of an eccentric planet still supports the latter as a better source of pollution.

\subsection{Non-gravitational forces}\label{sec:nongrav}
Post-MS evolution can have further implications for the orbits of small bodies due to non-gravitational effects. Wind drag during the giant branches can hinder the ability of small bodies to move outward, causing them to move relative to the ECZ and MMRs of a planet \citep{Dong:2010lr, Jura:2008ai}. Furthermore, the luminosity of the star increases dramatically during the giant phases. This brightening can produce a non-negligible Yarkovsky force on some small bodies \citep{Bottke:2006ve}, moving them inwards or outwards.  Both of these effects result in rearrangement of previously stable particles, repopulating some unstable regions and resulting in a new source of WD pollution. 

The wind resulting from stellar mass loss can strongly affect small bodies, particularly near the star. \citet{Dong:2010lr} examined this effect on objects orbiting more massive stars (3--4 M$_\odot$) at greater distances ($>10$ au) than our simulations, and found that wind drag can change the final orbit of even moderately-sized bodies (1--10 km) as well as cause resonance capture for a range of initial conditions. Using our own parameters, we analytically estimated the effect of wind drag on bodies located at 7 au around a 1 M$_\odot$ star. We found that at this distance 250-m diameter objects moved au-scale distances, and even 10-km diameter objects showed significant changes to their final orbit: roughly 0.04 au inwards compared  to the adiabatic case. Assuming a number distribution inversely proportional to size and a disc spanning 7 to 9 au, this wind drag results in approximately 1.5 per cent of the material entering the ECZ. While it depends on the width (due to larger objects moving less) and location of the disc, this order-of-magnitude calculation does indicate substantial amounts of mass can change location for large discs.

The Yarkovsky effect depends heavily on the physical properties of the small bodies, including size, shape, composition, and rotational state. However, by following prior research into the topic we estimated the change in SMA through this mechanism. According to \citet{Spitale:2001kx}, 100-m objects around the Sun have $\mathrm{d}a/\mathrm{d}t\sim 0.1$ km yr$^{-1}$ motion for most eccentricities. While changes to the spin states of the object reduce the long-term importance of the Yarkovsky effect during the MS \citep{Farinella:1998ys}, the brief period over which the giant branches last should result in few (if any) reorientations for objects 1 km in diameter or larger. Smaller objects will likely move rapidly, allowing for large changes in SMA even with reorientation. 

If we assume this motion scales inversely with the size of the object, the SMA of a 10-km body will then change by 0.0037 to 0.18 km yr$^{-1}$, or $2.4\times10^{-11}$ to $1.2\times10^{-9}$ au yr$^{-1}$, around an RGB star (average luminosity $\bar{L}\approx 180 L_\odot$), depending on the luminosity dependence of this effect: an asteroid with poor heat conduction will see the Yarkovsky effect increase linearly with the surface flux, but efficient heat conduction will reduce the temperature difference between the hot and cold faces and weaken the effect. We account for this uncertainty by considering a scaling of the strength of the effect with stellar luminosity ($L$) or with stellar effective temperature ($L^{0.25}$). Over the duration of the RGB branch, 80 Myr, the body would move a distance $\Delta a \approx $0.002--0.1 au.  While this estimate ignores the impact of mass loss (due to the complicated dependence on SMA), it nevertheless illustrates the power of the Yarkovsky effect during periods of high luminosity.

Both of these effects also impact the eccentricity of surviving bodies. The Yarkovsky effect is capable of pumping or damping it depending on the spin orientation of the body \citep{Spitale:2002uq}, while wind drag exclusively causes damping \citep{Dong:2010lr}. The latter affects eccentricity more strongly, causing the orbits of small bodies to become more circular during stellar evolution. Meanwhile, neither wind drag nor the Yarkovsky effect excite inclination (the latter as a result of orbital precession caused by the planet \citep{Bottke:2000sp}), leaving low-inclination bodies particles to continue on roughly coplanar orbits. 

Finally, it is important to note that the location and composition of the planet are crucial for its survival during stellar evolution. Planets near the star will be engulfed, while those just outside the envelope can still spiral in due to tidal effects \citep{Rasio:1996lr}. For massive planets farther out, the high flux received during and after the giant phases might result in planetary evaporation and mass loss \citep{Villaver:2007bh}.

\section{White dwarf simulations}\label{sec:WDsims}
To determine how stellar evolution affects the stability of orbiting TPs, we ran two sets of simulations. The first used the final orbital elements of the planet and surviving TPs from the MS simulations translated outward; the second used a new set of particles in the region around the WD.

\subsection{\emph{In situ} evolved simulations}\label{sec:evolved}
For the simulations that left particles in place during evolution, we first came up with a model for the evolution itself. We used a simple transition in which we reduced the mass of the star linearly over 2700 years, while leaving the planet and TPs \emph{in situ} with the orbital elements they had at the end of the MS simulations. We note that there are myriad ways to model mass loss in post-MS stars, including various durations and time dependencies. Additionally, a more accurate and complicated treatment would include the non-gravitational forces described in Section \ref{sec:nongrav}. However, this evolution is not our focus and as such we did not consider other, more-complicated models. Furthermore, the orbital expansion does not depend on the time-scale for mass loss, as long as the latter is much longer than the orbital periods (the adiabatic approximation). 

We implemented the mass loss in our code again using the BS integrator, with the stellar mass reduced by $5.61\times10^{-6}$ M$_\odot$ after each 12-day time-step. We chose the duration of the mass loss to minimize computation time while still being adiabatic, which is necessary for Equation \ref{eq:aexpand} to hold. In all the WD simulations the ejection and accretion distances remained the same at 100 au and 0.05 au, respectively. While the radius of a WD is dramatically smaller than that of a MS star, the Roche limit is approximately the same. 

We also mention that while our MS simulations did not reach the actual MS lifetime of a 1 M$_\odot$ star ($\sim 10^{10}$ years), the vast majority of unstable particles were removed by $10^8$ years (as illustrated by the number lost at late times in Figure \ref{fig:histlifetimes}). Using the change in loss rates over time, we estimate approximately 30 per cent of the unstable particles in these evolved systems were unstable in the MS system at times beyond $5\times10^{7}$ years. As such, we were able to approximately remove this contribution to the WD simulations and determine the instabilities introduced by mass loss. The 0.03 M$_\mathrm{Jup}$ simulations were an exception to this rule, due to longer particle lifetimes, and are addressed in Section \ref{sec:extsim}.

\subsubsection{Results}
\begin{figure*} 
\begin{center}
\includegraphics*[width=1.0\textwidth]{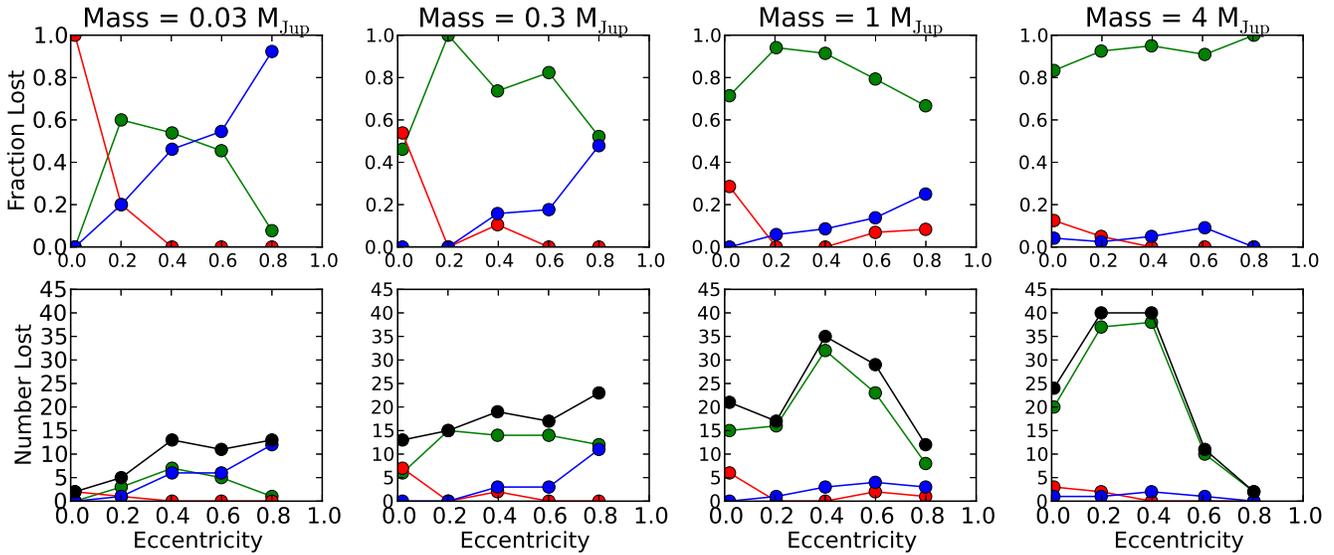}
\caption{Same plots as Figure \ref{fig:mseccfraclost}, fraction lost (top) and number lost (bottom) for each mechanism as a function of eccentricity, but for our evolved simulations. The eccentricity dependence still remains in most cases, but disappears in the highest-mass, highest-eccentricity cases due to the ECZ exceeding the bounds of the simulation. \label{fig:evolost}}
\end{center}
\end{figure*}

These simulations followed the same eccentricity--fraction-accreted relationship: increasing planetary eccentricity resulted in a larger percentage of unstable particles accreted by the star in all cases (Figure \ref{fig:evolost}). Unlike the MS simulations, however, this relationship did not correspond to a larger number of TPs accreted in every case. For the 1 and 4 $_\mathrm{Jup}$ planets at eccentricities of 0.6 and 0.8, the ECZ spanned nearly the entire 10 au disc in the MS, leading to fewer unstable particles in these simulations (Figure \ref{fig:evolost}, bottom right panels). Had the initial disc extended beyond 10 au initially, it is likely that the more-eccentric planets would have resulted in more unstable particles, similarly to planets with mass 0.03 and 0.3 M$_\mathrm{Jup}$ and as they did in the MS case. 

For the simulations with ECZs that did not span the entire disc during the MS, the total number of unstable particles varied more strongly with planetary mass than during the MS. As described in Section \ref{sec:evo}, the ECZ and MMRs widen as a result of mass loss, which produces newly unstable particles. Both the fractional growth in the ECZ (Equation \ref{eq:eczexp}) and absolute change in size depend on planetary mass. Therefore it is not surprising to see this stronger mass dependence on total number of unstable particles. We observed the ECZ expansion in the results of some simulations, but did not see it consistently between masses or eccentricities as a result of small number statistics, particularly in the higher eccentricity cases where few TPs survived from the MS.

 \subsubsection{Extended simulation}\label{sec:extsim}
Due to the late onset of instability in the MS 0.03 M$_\mathrm{Jup}$ case, we expected a significant fraction of unstable TPs to be unstable on time-scales greater than $10^8$ years, resulting in contamination to the evolved simulations.  To combat this issue and the low number of TPs noted above, we repeated the 0.03 M$_\mathrm{Jup}$, $e=0.4$ simulation with quintuple the TP resolution (reducing the separation between TPs to 0.004 au) and for a longer duration: $2\times10^8$ years for the MS and $1\times10^9$ years for the WD phase. Long computation times prevented us from repeating this simulation for additional planetary masses and eccentricities. Such a high density  of TPs remedied the issue of few survivors in the post-evolution simulation, allowing us to better test the potential of a single, ideal planet to pollute a WD without non-gravitational forces. 

The MS simulation behaved identically to the same planet in Section \ref{sec:msresults}, with the exception of better statistics and an increased duration. The lost particles peaked near $5\times10^{6}$ years, and accretion dominated the unstable particles while collisions with the planet occurred infrequently. The extra 100 Myr behaved similar to the prior 50 Myr, showing both accretion events and ejections at a much reduced rate from the peak. Extrapolating from the accretion rate as a function of time, we found that that only $\sim25$ particles were expected to go unstable in the subsequent 1 Gyr, which allowed us to account for contamination in our next simulation. Finally, as the number of TPs were not a limiting factor in the prior MS simulation, these results do not gain us much more insight. 

 \begin{figure} 
\begin{center}
\includegraphics*[width=0.5\textwidth]{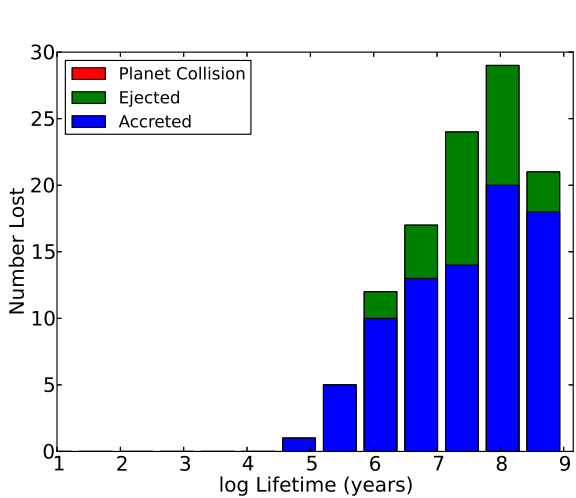}
\caption{The lifetimes of unstable particles for the evolved system with  0.03 M$_\mathrm{Jup}$, $e=0.4$, illustrating the late onset of instability and the large fraction of accreted particles. \label{fig:exthist}}
\end{center}
\end{figure}

\begin{figure}
\begin{center}
\includegraphics[width=.5\textwidth]{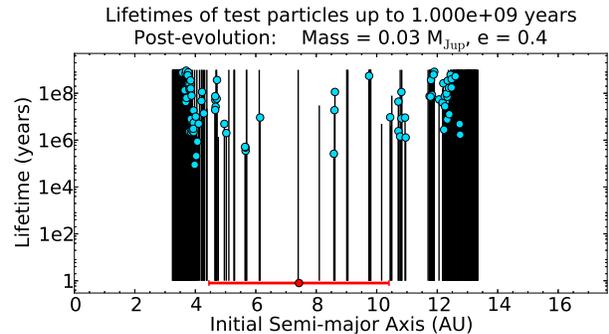}
\caption{Lifetimes of TPs for the evolved extended simulation (0.03 M$_\mathrm{Jup}$ and $e=0.4$, initial spacing of 0.004 au). The expansion of the ECZ due to stellar mass loss is apparent as is the large fraction of accreted particles. \label{fig:extendlifetimes}}
\end{center}
\end{figure}

The evolved simulation, however, differed significantly: the increased number of surviving particles and greater duration produced an obvious expansion of the ECZ (Figure \ref{fig:extendlifetimes}). The ECZ expanded by 10 per cent, less than expected in the circular case but larger than expected from Equation \ref{eq:ECZ}, 3.1 per cent. This simulation also displayed the accretion and ejection peaking near $10^8$ years, as shown in Figure \ref{fig:exthist}, with the 20-fold increase from MS peak due to both fewer particles unstable on short time-scales (cleared out in the MS) and the larger planetary SMA. Using the MS-simulation extrapolation, we found that that fewer than 25 per cent of lost particles were contaminants. Significant accretion continued to occur up to the end of the simulation, indicating that a planetary system like this, given certain disc properties, could account for the pollution observed in some polluted WDs. We discuss the feasibility of this mechanism in further detail in Section \ref{sec:modelfit}. 

\subsection{Repopulated simulations}\label{sec:repopulated}
The second set of WD simulations represented complete repopulation by non-gravitation forces, and we populated these with a completely new set of TPs spaced 0.03 au apart\footnote{The spacing was increased from 0.02 au to allow us to probe the larger spatial scales with similar computation time.} between 0.06 au and 18.6 au. For consistency with the MS simulations and the effects described in Section \ref{sec:nongrav}, all particles started at random points on circular orbits. These simulations were similar to the original MS simulations, but with all bodies at greater distances about a reduced stellar mass. As a result, they represented the change in stability of locations within the system, as opposed to the change in stability of individual particles (which showed some small amounts of motion within stable regions during the MS). While complete repopulation of all unstable regions in a system is unrealistic, as much as 1.5 per cent of the mass in a narrow annulus can be expected to move into them during stellar evolution via non-gravitational forces (as described in Section \ref{sec:nongrav}). Alternatively, planetary systems that are near instability would result in a planet ending up on a new orbit after a scattering; the results of  \citet{Veras:2013vn} indicate that such scatterings are possible around WDs at late times, and can result in highly-eccentric orbits.

\subsubsection{Results}\label{sec:ECZevo}

\begin{figure*}
\begin{center}
\includegraphics*[width=1\textwidth]{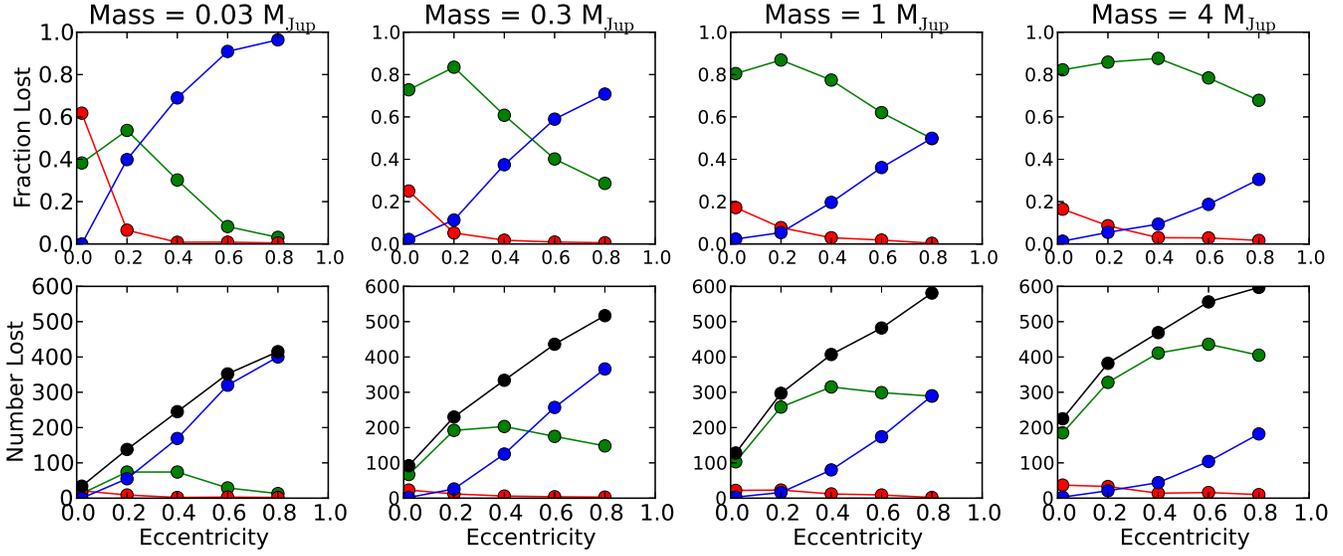}
\caption{Same plots as Figure \ref{fig:mseccfraclost}, fraction lost (top) and number lost (bottom) for each mechanism as a function of eccentricity, but for our repopulated WD simulations. The same characteristics as in the MS simulations are present (e.g. increase in accretion fraction with planetary eccentricity) but the ejection fraction is systematically higher in all cases.\label{fig:wdfraclost}}
\end{center}
\end{figure*}

As illustrated in Figure \ref{fig:wdfraclost}, these simulations behaved similarly to those of Section \ref{sec:msresults}, . Increased eccentricity again dramatically increased both the total number of TPs that went unstable as well as the fraction that were accreted by the star, and lower masses again showed equivalent or greater numbers of accreted particles. We noted certain differences between the two sets of simulations, however. Due to the increased distance from the star and the reduced stellar mass, TPs were more weakly bound and a greater fraction ejected in the WD setup. This change can be seen in a comparison of Figures \ref{fig:mseccfraclost} and \ref{fig:wdfraclost}, and correspondingly led to a smaller accretion fraction. 

Compared to the extended simulation, the repopulated 0.03 M$_\mathrm{Jup}$, $e=0.4$ simulation peaked earlier and had a smaller unstable fraction at late times due to the large contribution from TPs in repopulated regions, which rapidly went unstable. Even so, both the repopulated and the extended models showed agreement in the fraction of particles accreted and ejected, allowing us to extrapolate accretion rates to other masses and eccentricities from our single extended simulation. Additionally, the location of the oldest particles remained the same:  the edge of the ECZ.
 
\begin{figure}
\begin{center}
\includegraphics[width=0.5\textwidth]{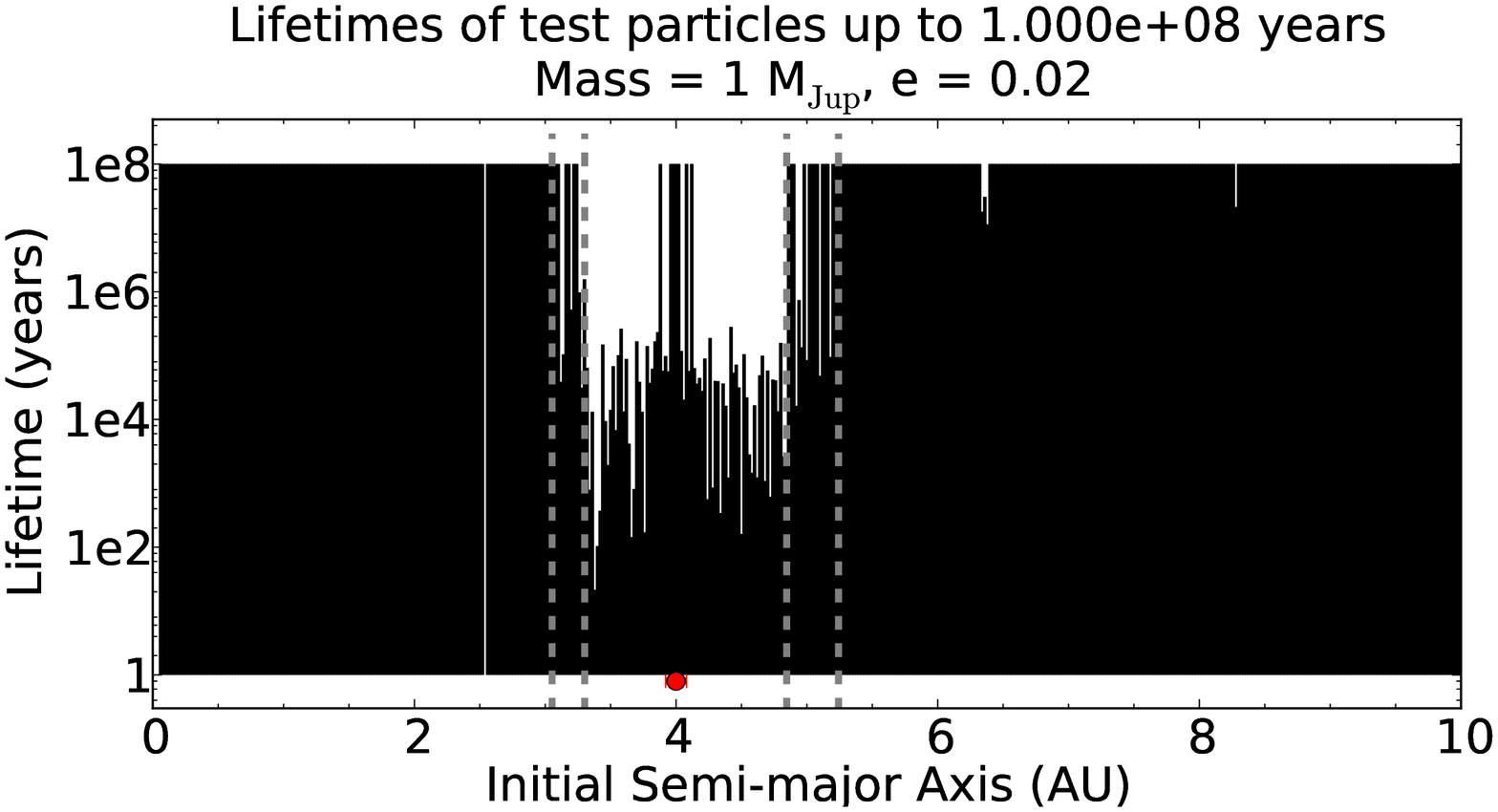}\\
\includegraphics[width=0.5\textwidth]{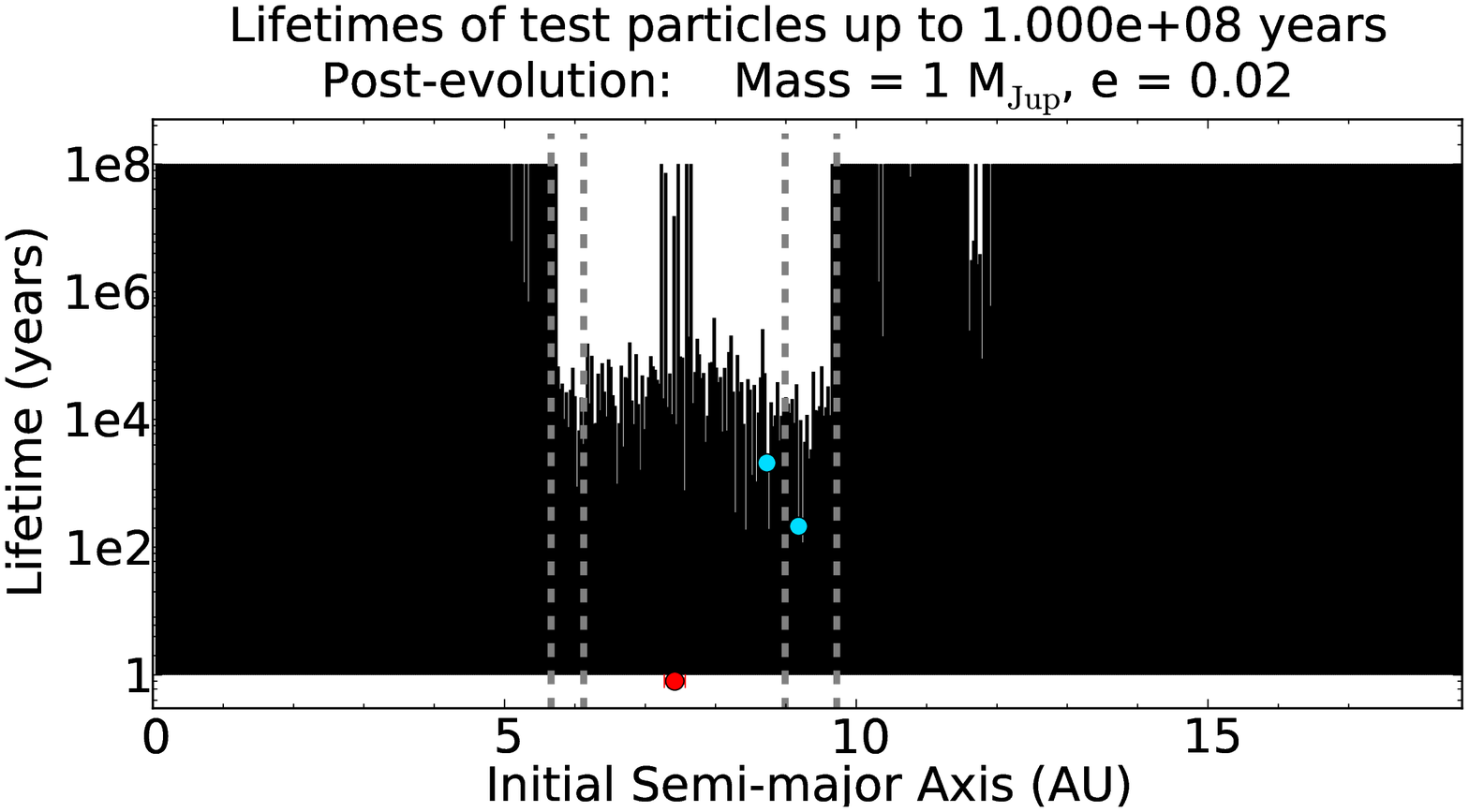}
\caption{A comparison of the MS and WD simulations for a planet with M = 1 M$_\mathrm{Jup}$ and $e=0.02$. Note the wider instability in the WD case, located between the 3:2 and 2:1 resonances and between the 1:2 and 2:3 resonances (grey dashed lines).\label{fig:unstabincrease}}
\end{center}
\end{figure}

\begin{figure}
\begin{center}
\includegraphics*[width=0.5\textwidth]{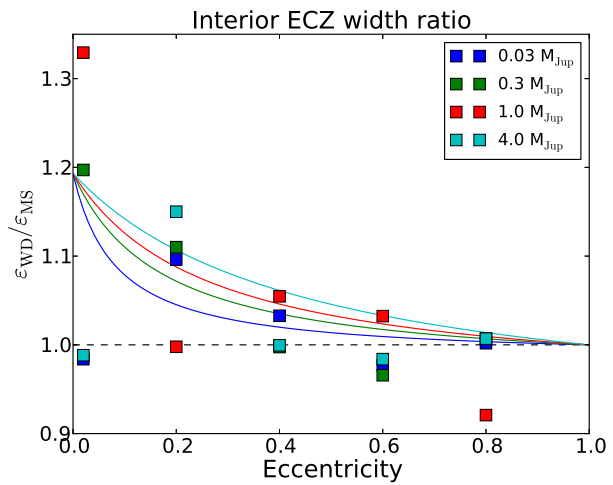}
\caption{Ratio of WD ECZ size to MS ECZ size, for the interior edge, with lines indicating no change in size (black dashed) and a change characterized by Equation \ref{eq:ECZ}(coloured solid). Despite the scatter, the trend of less expansion with greater eccentricity is clear.\label{fig:ECZgrowth}}
\end{center}
\end{figure}

As a result of stellar mass loss and orbital expansion, the width of the ECZ should increase according to Equation \ref{eq:ECZ}. We observed this expansion in our repopulated simulations, illustrated in Figure \ref{fig:unstabincrease} for a near-circular planet. The sizes of the ECZ were similar to that shown in Figure \ref{fig:ECZwidth}, and are not plotted. As described in Section \ref{sec:evo}, the relative increase in the size of the ECZ should decrease with eccentricity when it obeys Equation \ref{eq:ECZ}. We found that to be generally the case for the interior edge of the ECZ, which followed Equation \ref{eq:ECZ}  closely in both the MS and the WD cases. Figure \ref{fig:ECZgrowth} illustrates this effect, as larger eccentricities generally have ratios closer to one. Conversely, the exterior edge deviated more so from Equation \ref{eq:ECZ} in both cases, and thus displayed greater scatter about the expansion relation. Additionally, the full size of the ECZ may be larger than plotted, particularly for the low mass simulations, as our extended simulation showed further particles becoming unstable at the ECZ edge after $10^8$ years.

Similar to the ECZ, the widths of MMRs should grow with mass loss and thus can be a valuable source of unstable particles \citep{Debes2012}. Our results showed this widening of MMRs in some of our simulations, particularly in the case of MMRs interior to massive planets. The 3:1 resonance in our $e=0.2$, 4 M$_\mathrm{Jup}$ simulation grew from 0.064 au in the MS case to 0.141 au in the WD case. While the physical width was expected to increase due to the  larger spacing between bodies, that effect would only result in a factor of M$_\mathrm{MS}/$M$_\mathrm{WD} = 1.85$, not 2.2 as we saw. In both cases a large fraction of the unstable particles were accreted (9 out of 16 for MS, 9 out of 19 for WD), despite a low fraction (13 per cent for both) of interior TPs accreted overall. Despite starting with fewer TPs than \citet{Debes2012}, these simulations support internal MMRs as another potential source of WD pollution. Given the results of our extended simulation in Section \ref{sec:extsim}, multiple mechanisms may play a role in the pollution observed.

 \subsubsection{Accretion at late times}\label{sec:latetimeaccr}
\begin{figure}
\begin{center}
\includegraphics*[width=0.5\textwidth]{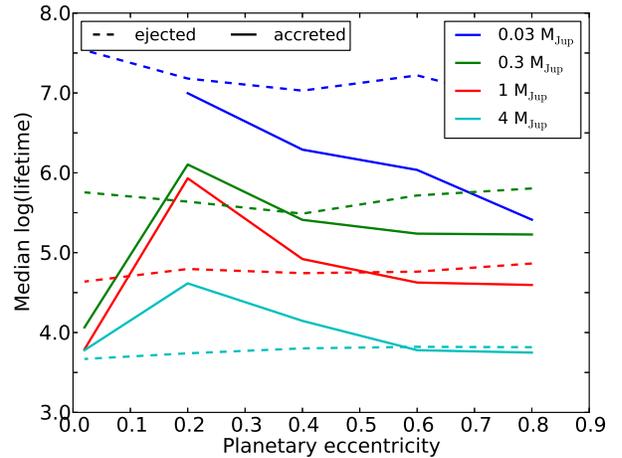}
\caption{Median lifetime of ejected and accreted unstable particles in log space for each planetary mass as a function of eccentricity, for our repopulated WD simulations. Very few particles were accreted in all the $e=0.02$ simulations (including zero for the $0.03$ M$_\mathrm{Jup}$ case), which accounts for the strange behavior and missing point at that eccentricity. All of the planets show in a decrease in accreted-TP lifetime when the planetary eccentricity is increased above 0.2. \label{fig:WDlosstimesplot}}
\end{center}
\end{figure}

\begin{figure}
\begin{center}
\includegraphics*[width=0.5\textwidth]{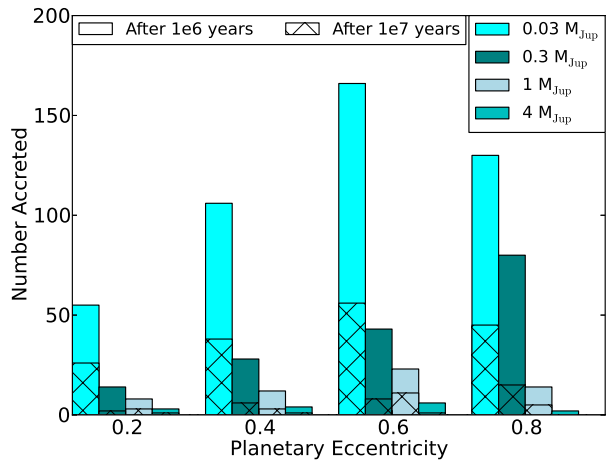}
\caption{Number of particles accreted by the WD in the repopulated simulations after $10^6$ and $10^7$ years, as a function of mass and eccentricity ($e= 0.02$ not shown, as nearly no particles were accreted in that case). The accretion amounts clearly increase with eccentricity for both times through $e=0.6$, across all masses, and decreases with mass across all eccentricities. \label{fig:WDlatetimes}}
\end{center}
\end{figure}

While the relationship between planetary properties and the total number of particles accreted over all time is interesting, it has little bearing on WD pollution unless that accretion occurs at late times. The results from our WD simulations agreed with our findings in Sections \ref{sec:MSmass} and \ref{sec:ecc} regarding particle lifetime: both higher masses and higher eccentricities lead to earlier peak times, as shown in Figure \ref{fig:WDlosstimesplot}. Because of this shift in peak time, the maximum amount of accreted material at late times (above $10^6$ and $10^7$ years) did not always occur in the simulations with the largest eccentricity, as shown in Figure \ref{fig:WDlatetimes}. Larger eccentricities increased the pollution at late times only up to a point, above which the pollution was reduced.  These peak eccentricities depended on planetary mass: $e=0.6$ in the case of 0.03 and 4 M$_\mathrm{Jup}$, and $e=0.8$ in the case of 0.3 and 1 M$_\mathrm{Jup}$. Given such large peak eccentricities, we confirm  our previous results: in the vast majority of cases, larger planetary eccentricities (and smaller masses)  correspond to larger amounts of polluting material. 
 
\section{Discussion}\label{sec:disc}
These simulations clearly show that planetary mass and eccentricity play an important role in the ability of a planet to pollute the central star. Given the mass dependence of both the fraction and the total number of accreted TPs, it appears that planets do not need to be as massive as Jupiter to be a potential source of WD pollution, and such massive planets may in fact not be the prime candidates. The extended simulation of Section \ref{sec:extsim} supported the result that small planets deliver as much material or more than massive planets, and at later times, with a 0.03 M$_\mathrm{Jup}$ planet continuing to deliver material at Gyr time-scales. It would not be unusual for an exoplanet to have these properties: small planets now appear to be more common than large planets, both near the star \citep{Batalha:2013yq} and further away \citep{Gould:2006lr, Sumi:2010fk}. Furthermore, a wide range of eccentricities have been detected in exoplanet surveys \citep{Butler:2006yq}, up to 0.4 for 0.03 M$_\mathrm{Jup}$ planets and above 0.8 for 4 M$_\mathrm{Jup}$ planets. The planetary mass which produces maximum accretion is still uncertain--- the smallest of our planetary masses (0.03 M$_\mathrm{Jup}$) produced the highest number of accretion events, but it is possible that even smaller masses would produce a higher number. However, it is not clear theoretically or observationally whether sub-Neptune-mass planets can form at several au distances. Such simulations would also be more computationally expensive, and possibly violate the massless-TP assumption.

\subsection{The WD disc mass from accretion rates}\label{sec:modelfit}
While our simulations did not reach the times corresponding to the oldest cooling ages of polluted WD ($>10^9$ years, \citet{Farihi:2010fj}), our longest simulation did reach the time-scales of average polluted WDs ($10^8$--$10^9$ years). Determining which planetary properties best cause pollution in a star is only useful if such pollution can match observations, which is set by the disc mass required in our simulations. To determine that mass, we assumed a power law for the loss rate of particles:

\begin{equation}
\frac{\mathrm{d}N_\mathrm{lost}}{\mathrm{d}t} = -\frac{\mathrm{d}N_\mathrm{rem}}{\mathrm{d}t}=\frac{N_\mathrm{rem}^{\alpha}}{t_0}
\end{equation}
Here $\alpha$ and $t_0$ are constants to be fit by our simulation results, and $N_\mathrm{rem}$ ($N_\mathrm{lost}$) is the number of particles remaining (lost). 
Solving this equation for $N_\mathrm{rem}$ gives
\begin{equation}
N_\mathrm{rem} = \frac{N_0}{[1 + (\alpha - 1)N_0^{\alpha-1}t/t_0]^{\frac{1}{\alpha -1}}} =  \frac{N_0}{[1 +t/t_*]^{\frac{1}{\alpha -1}}}
\end{equation}
where $N_0$ is the initial number of unstable particles ($N_\mathrm{rem} = N_0 -  N_\mathrm{lost}$), another constant to be fit, and
\begin{equation} 
t_*= \frac{t_0}{(\alpha - 1)N_{0}^{\alpha-1}}
\end{equation}
is the characteristic time scale for losing particles. The loss rate as a function of time is then
\begin{equation}\label{eq:dndt}
\frac{\mathrm{d}N_\mathrm{lost}}{\mathrm{d}t} =-\frac{\mathrm{d}N_\mathrm{rem}}{\mathrm{d}t} = \frac{N_0}{(\alpha - 1)t_*}(1 + t/t_*)^{\frac{-\alpha}{\alpha - 1}}
\end{equation}

\begin{figure} 
\begin{center}
\includegraphics*[width=0.5\textwidth]{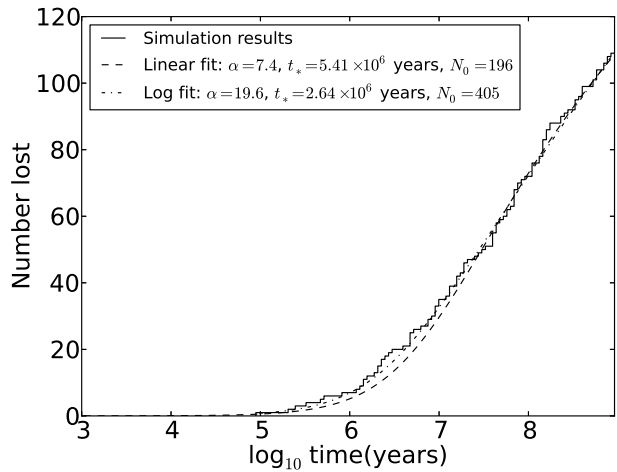}\\
\includegraphics*[width=0.5\textwidth]{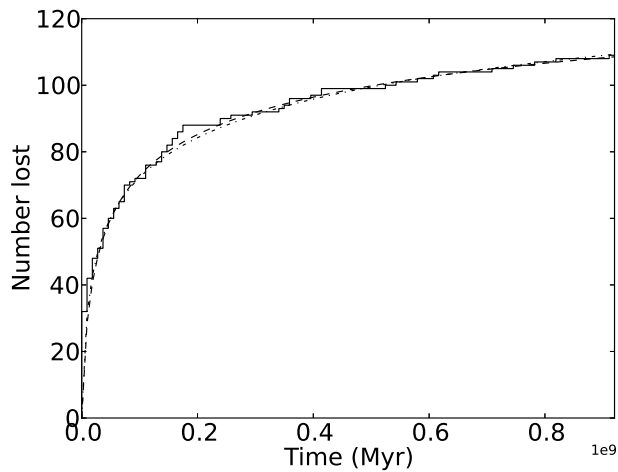}
\caption{The cumulative number of particles lost through all mechanisms, as a function of time in log$_{10}$ space (top) and linear space (bottom), with the power law fit to the log (linear) distribution shown in as a dash-dotted (dashed) line. These fits allowed us to specify the loss rate at 1 Gyr and beyond, which we compared to accretion rates around WDs to produce an estimate of the initial unstable mass.\label{fig:cumplots}}
\end{center}
\end{figure}

We determined the values of $\alpha$, $t_*$, and $N_0$ by performing a least-squares fit to the cumulative distribution of lost particles. We selected the evolved extended run (M$=$ 0.03 M$_\mathrm{Jup}$, $e=0.4$) for fitting due to the larger number of TPs and extended duration. This simulation defined the `best-case' planet, as higher masses resulted in shorter TP lifetimes and greater ejection fractions while larger eccentricities have yet to be observed above 0.4 for low-mass planets. The best-fitting values were $\alpha=7.4$, $t_*=5.41\times10^6$ years, and $N_0 = 196$ particles. We also fit the cumulative distribution with bins in $\log_{10}$ space, which gave different values: $\alpha=19.6$, $t_*=2.64\times10^6$ years, and $N_0 = 405$ particles. The data and the fits are shown in Figure \ref{fig:cumplots}. Although the two sets of fitted constants differed greatly, the fits themselves produced similar fractions of material accreted, differing by less than a factor of two at any given time, allowing either fit to work in an estimate of the total unstable disc mass.

For the linear-fit values and a time of $10^9$ years, Equation \ref{eq:dndt} gives $1.35\times10^{-9}$ particles removed per year. Dividing by $N_0$ gives the fractional loss rate, $6.87\times10^{-11}$ per year. We assume that the 25 per cent contamination from MS unstable material (Section \ref{sec:extsim}) affects both the particle loss rate and $N_0$ equally, so the fraction is unaffected. Examining each loss mechanism individually as a function of time (Figure \ref{fig:exthist}), we find that the accretion rate equals roughly double the ejection rate after $10^7$ years, and the planetary collision rate is negligible. Therefore we assume the accretion rate is two-thirds of the total loss rate, or $\approx4.58\times10^{-11}$ of the initial material per year.  We can compare this to the observed accretion rate of polluted WDs, roughly $10^{8}$ g s$^{-1}$ for WDs with cooling ages of $10^9$ years \citep{Farihi:2010fj}, which gives an initial mass of unstable bodies in this system equivalent to $6.9\times10^{25}$ g. This mass is roughly 23 times that of the asteroid belt, $\sim 3\times10^{24}$ g \citep{Pitjeva:2005mi}, and one per cent the mass of the Earth (0.01 M$_\oplus$). While larger than previous estimates of the total accreted material \citep{Zuckerman:2010kl}, this mass is negligible compared to our smallest planet and supports the use of massless TPs.

Using the results from Section \ref{sec:repopulated}, we repeated the calculation for our repopulated simulation with the same mass and eccentricity and found a larger required disc mass: the material in the repopulated regions reduced the fraction accreted at late times and thus the total amount was necessarily larger. More-massive and lower-eccentricity planets, as expected, required even greater initial disc masses, both due to the shorter lifetimes of small bodies (in the case of massive planets) and due to the reduced fraction of accreted material relative to ejected material. Additionally, non-gravitational forces have only a minor effect on the required mass, due to the limited amount of motion provided to large bodies. As such, we conclude that for a planet orbiting at 7.42 au about a 0.539 M$_\odot$ star with no external companions, an accompanying disc must have at least 0.01 M$_\oplus$ in unstable material in order to account for the observed levels of WD pollution even in the best case (lowest mass and highest eccentricity) scenario.

\subsection{Observational and dynamical disc constraints}\label{sec:maxmass}
To determine whether this result is reasonable in the context of observed debris-disc masses, we extrapolate the total disc mass from the amount of unstable material. To do so, we assume that the unstable material is localized to a region between 11.7 and 12.7 au, beyond which the objects are stable. We make this assumption based on the results of our extended-duration simulation (Section \ref{sec:extsim}), in which the majority of unstable particles exterior to the planet were located in that region. We also assume the disc extends to 90 au (expanded from $50$ au, in analogy with the edge of our solar system \citep{Trujillo:2001kx}) and has a surface density of the form $\Sigma(r) = \Sigma_0 (r/r_0)^{-3/2}$ \citep{Kenyon:2004uq}. Integrating over the unstable part of the disc and equating that to the result of Section \ref{sec:modelfit} allows us to determine $\Sigma_0$, from which we can calculate the total disc mass. Doing so, we find that the total mass, in both stable and unstable regions, is 0.5 M$_\oplus$, 50 times the unstable mass. This value is strongly dependent on the size of the disc as well as the presence of any other planets, which would carve out additional ECZs and reduce the amount of material remaining at the end of the MS.

This mass is not unreasonable, as observed debris discs show dust masses of $10^{-2}-10^{-1}$ M$_\oplus$ around stars older than 1 Gyr \citep{Wyatt:2008th}. Due to the fact that large bodies dominate the mass while small bodies dominate the surface area for most  planetesimal size distributions \citep{Wyatt:2007wd}, the total mass of these observed discs can be as much as $10^{3}-10^{5}$ times larger \citep{Lohne:2008kx}, well above our requirement. Furthermore, while this mass is much larger than the asteroid belt, the latter may have been substantially thinned out by the peculiarities of  Solar System evolution (e.g. \citet{Walsh:2011fk}), making such a comparison unnecessarily restrictive. Minimum mass solar nebula estimates yield original masses $\sim 100$ times larger than the present value \citep{Weidenschilling:1977qy}, more than enough to meet pollution requirements if concentrated in a narrow region similar to the current asteroid belt.

However, while massive discs are not uncommon around MS stars, the question of whether they survive near the star and at late times arises, due to the effect of collisional evolution and radiative forces depleting much of the material over the stellar lifetime. Assuming an infinite collisional cascade, \citet{Wyatt:2007wd} show that such forces result in a maximum disc mass for a given age. This maximum depends on the physical properties of the disc such as width, distance from the star, and the properties of the constituent particles. 

\begin{figure}
\begin{center}
\includegraphics*[width=0.5\textwidth]{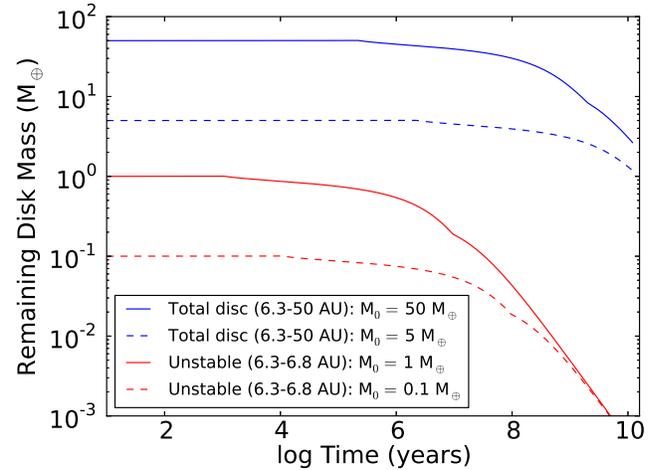}
\caption{Evolution of disc mass as a function of time, for the total mass (blue) and the unstable mass (red). While the total mass is large enough to match the mass required by WD accretion rates, the unstable mass is over an order-of-magnitude smaller than our requirement (0.01 M$_\oplus$). \label{fig:discevo}}
\end{center}
\end{figure}

To determine if our required mass can exist in a disc 10 Gyr old, we calculated the maximum disc mass for both the unstable region and the total disc using the \citet{Wyatt:2007wd} model. At the end of MS, the material will not have moved outward due to mass loss, so the unstable region will span 6.3 to 6.8 au with the entire disc reaching 50 au. Using these disc dimensions, a largest object size of 2000 km, and an assumed particle eccentricity of 0.25 (the forced eccentricity at 6.5 au), we find that the maximum amounts of unstable and total material are $6\times10^{-5}$ M$_\oplus$ and $0.16$ M$_\oplus$, respectively. While our parameter selections can affect these values and increase the maximum total mass above our 0.5 M$_\oplus$ requirement, the amount of unstable material is dramatically smaller than what needs to exist to account for WD pollution regardless.

The assumption of collision equilibrium holding for all mass sizes has been challenged by \citet{Lohne:2008kx}, who argue that it is valid only at very late ages and show that massive discs can exist even at 10 Gyr for a range of parameters. We repeated the mass-remaining calculation using their formalism and the same disc parameters, and found that although the total disc mass was relatively unconstrained, the mass in the unstable region was limited to $10^{-3}$ M$_\oplus$. These results (shown in Figure \ref{fig:discevo}) indicate that, while the disc mass can reach larger values at 10 Gyr than in the \citet{Wyatt:2007wd} case, the masses still fall short of what is required to match observed pollution rates when the disc is near the star. Therefore, we determine that collisional evolution prevents a planet with a narrow unstable region from being a major source of WD pollution, unless it is significantly more distant than 4 au during the MS or the star loses a significantly larger mass fraction.  Alternatively, if collisional evolution of the material progresses differently from what current models predict, the original reservoirs could be massive enough to supply polluting material, depending on the planetary properties. We should mention that this collisional evolution may prevent the MMR-based mechanism discussed in \citet{Debes2012} from being a viable source of pollution as well, due to the massive asteroid belt required (0.35 M$_\oplus$) in their simulations.

\subsection{Other system parameters}
While we examined the effects of some planetary-system properties on WD accretion rates in detail, it is useful to consider the impact of other parameters as well. One such parameter is the initial SMA of the planet, which remained constant at 4 au between all our cases. We ran a single simulation to investigate this effect, using a 0.03 M$_\mathrm{Jup}$, $e=0.4$ planet with an SMA of 10 au. We found a small decrease in the fraction of particles accreted (from 83 per cent to 78 per cent) but a substantial increase in the fraction of TPs accreted in the last 50 Myr of the simulation, resulting from the increased orbital time-scale. That fraction increased from 3 per cent in the 4 au case to 8 per cent in the 10 au case, indicating that a smaller disc mass (both total and unstable) may be required for planets at larger distances. 

Farther from the star the maximum disc mass should be greater, potentially allowing a single planet to be a source of WD pollution. As a test, we repeated our calculations to determine the maximum unstable mass in a disc scaled outward by a factor of 2.5. We found that, in a disc located from 15.75 to 17 au with the same eccentricity and maximum object sizes as before, the \citet{Wyatt:2007wd} and \citet{Lohne:2008kx} approaches predict $10^{-3}$ M$_\oplus$ and 0.015 M$_\oplus$, respectively. Given the potentially-lower disc-mass requirement due to increased late-time accretion, these results indicate that more distant planets and discs may serve as a more-likely source of pollution.

The initial stellar mass also plays a major role in system dynamics, determining the orbital expansion experienced by the planet and other bodies. More-massive stars undergo greater mass loss, resulting in a larger expansion and potentially longer lifetimes as described above. The results of \citet{Wisdom1980} predict that this larger expansion would have a correspondingly larger increase in the unstable region, which could result in a larger population of small bodies accreting on to the star. Furthermore, more-massive stars have shorter lifetimes, allowing them to retain more planetesimals. However, the larger separation from the central star would likely reduce the accretion fraction relative to our results, as we saw in the lower fraction for TPs around the WD compared to the MS star. 

Doubling the stellar mass for some MS simulations, we saw a small increase in the fraction of particles accreted along with a decrease in ECZ size, resulting in fewer total accreted. As a result of the shorter planetary period, the average lifetime of TPs shrank and thus even fewer particles accreted after $10^7$ years. While we did not simulate the evolved case of this increased stellar mass, we can assume, from the two prior results, that the particle lifetimes would be longer and the accretion fraction less. The magnitude of the accretion reduction would depend on the total mass loss of the star, and therefore the initial stellar mass.

\section{Conclusion}\label{sec:conc}
In this work we have simulated single-planet systems through stellar evolution, allowing us to detail the effect of planetary parameters on the WD accretion rate and determine that, when started in a disc initially at zero eccentricity, planets of mass $\le0.03$ M$_\mathrm{Jup}$ and eccentricity $e\ge0.4$ are the most-efficient perturbers. We find that more-massive planets deliver less material to their host star due to ejecting a much larger fraction of unstable particles, while smaller eccentricities also produce a lower accretion fraction in addition to fewer unstable particles. The mass in planetesimals, while found to be negligible for Jupiter-mass planets, can have dynamical consequences if the mass of the scatterer is too small; the use of massless TPs in simulations thus limits our results to planetary scatterers larger than several Earth masses. Particle lifetime varies inversely with both planetary mass and eccentricity, but the latter relationship is weak relative to the variation in overall accretion amount. These relationships remain for non-zero particle eccentricity, but are significantly weakened when particle longitude of pericentre matches that of the planet.

We further find that stellar evolution has an impact as well, widening planetary and particle orbits thus causing an increase in the fraction of particles ejected and a decrease in the overall accretion amount. Longer orbital periods translate into later peak accretion times, which partially offset the latter effect. Additionally, while non-gravitational forces become stronger during stellar giant phases, they play a negligible role in the amount of stellar accretion due to strong dependence on planetesimal size.

Most importantly, we have demonstrated that single-planet systems within 8 au of their host WD can deliver the amount of material observed in polluted-WD atmospheres through the scattering of small bodies from reservoirs similar in mass to that which existed early in our own planetary system. These small bodies were assumed to be at random points on circular orbits at the beginning of the MS. For a low-mass, high-eccentricity planet (0.03 M$_\mathrm{Jup}$, $e=0.4$) at an SMA of 7.42 au, $10^{-2}$ M$_\oplus$ of unstable material would be required in a disc annulus 1 au wide, which is within observational amounts. However, current models of collisional evolution predict that the accompanying disc cannot retain an adequate reservoir of material, as much of it is ground down to dust and lost from the system. If these collisional models are correct, more distant planets (where disc evolution progresses more slowly) or planets that are scattered to other portions of the disc (as described in \citet{DebesSigurdsson}) remain a possible source of pollution.

We finish by mentioning that, so far, no planets have been discovered orbiting a WD, regardless of metal pollution or infrared excess \citep{Faedi:2011ys, Hogan2009}. This is unsurprising in the context of our results, given the low planetary mass necessary to destabilize small bodies and the current difficulty in detecting exoplanets around any star at 7 au. With radial-velocity detections limited to greater than $\sim1$ km s$^{-1}$ due to pressure broadening of spectral lines \citep{Maxted:2000ly}, even massive planets are unlikely to be detected in that manner. If planets such as super-Earths and Neptunes are significantly more common than more-massive planets as sources of pollution, then it would further reduce the possibility of detecting them in the future through methods such as direct detection and astrometry. 

\section*{Acknowledgments}
We thank Ian Crossfield, Julia Fang, and Siyi Xu for their helpful comments, as well as the anonymous referee for their comments, which greatly improved the clarity and conclusions of this paper.

\bibliographystyle{mn2e}   
%\bibliography{/Users/sfrewen/Research/research}

\label{lastpage}

\end{document}